\documentclass[letterpaper,twocolumn,pre,
aps,showpacs,superscriptaddress,
floatfix]{revtex4-2}

\usepackage{tikz}
\usepackage{amssymb,amsmath}
\usepackage{graphicx}
\usepackage{booktabs}
\usepackage{verbatim}
\usepackage{lmodern} 
\usepackage{hyperref}
\usepackage{url}
\usepackage{csquotes}
\usepackage{epstopdf}
\usepackage{booktabs}
\usepackage{verbatim}
\usepackage[IL2]{fontenc}  
\usepackage[caption=false]{subfig}
\usepackage{algpseudocode}
\usepackage{algorithm}

\usepackage{array}

\begin{document}

\begin{abstract}
We report a dynamical phase transition in the information spreading within a classical 2D deterministic interacting many-body system. Specifically, the transition is observed in a recently introduced momentum-conserving parity check cellular automaton (MCPCA) on the square lattice~\cite{Kasim2025}. We characterize the transition using information-theoretic quantities such as the Hamming distance and the classical decorrelator. By introducing conserved local charges of the MCPCA, we show that selecting initial ensembles with specific charge values allows the system to transition from a localized information phase to a chaotic regime with ballistic information spreading. Importantly, our findings indicate that this transition is of second order, highlighting a sharp change in information spreading behavior. Furthermore, we revisit the multifractal behavior of the dynamical structure factor and show that, although present across both phases, it originates from effective local periodicities enforced by symmetry constraints.

\end{abstract}

\title{Phase transition from localization to chaos in classical many-body system}

\author{Yusuf Kasim}
\affiliation{Faculty of Mathematics and Physics, University of Ljubljana, Jadranska 19, SI-1000 Ljubljana, Slovenia}
\author{Pavel Orlov}
\affiliation{Faculty of Mathematics and Physics, University of Ljubljana, Jadranska 19, SI-1000 Ljubljana, Slovenia}
\affiliation{Nanocenter CENN, Jamova 39, SI-1000 Ljubljana, Slovenia}
\author{Toma\v{z} Prosen}
\affiliation{Faculty of Mathematics and Physics, University of Ljubljana, Jadranska 19, SI-1000 Ljubljana, Slovenia}
\affiliation{Institute of Mathematics, Physics and Mechanics, Jadranska 19, SI-1000 Ljubljana, Slovenia}
\date{\today}
\maketitle

\section{Introduction} 

The regularity-to-chaos transition is one of the most ubiquitous transitions in nature. In few-particle classical systems, this transition is well understood through the celebrated Kolmogorov–Arnold–Moser (KAM) theory~\cite{ArnoldBook}. 
For interacting many-body systems, however, our understanding remains incomplete once the coupling strength -- or equivalently, the integrability-breaking parameter -- pushes the system beyond the range of perturbation theory. In this regime, fundamental phenomena emerge, including the onset of ergodicity and chaos, thermalization, and the crossover from ballistic to diffusive transport. This continues to be a highly active field of research, with very few exact or rigorous results despite more than half a century of effort. A historic testbed for this type of research has
been the Fermi-Pasta-Ulam-Tsingou model~\cite{Fermi55,Izrailev}. 
One of the central questions concerns how the critical coupling strength scales with system size in the thermodynamic limit. Although it is widely believed that this strength typically vanishes even for local interparticle interactions in low dimensions (in both classical and quantum settings), there are essentially no exact or rigorous results. Numerical experiments even suggest alternative possibilities, see e.g.~\cite{Prosen98,Prosen07}.
A major open challenge is to identify minimal, exactly solvable many-body models that exhibit the transition from regularity to chaos or ergodicity. In this work, we present such a candidate system.

A popular measure of dynamical complexity—and a hallmark of quantum chaos in quantum systems—is the out-of-time-order correlator~\cite{Maldacena2016,Hashimoto2017, Swingle:2018ekw}. Its classical counterpart, the decorrelator, has recently been employed to probe the spread of classical information and to quantify the butterfly velocity~\cite{Bilitewski2018,Das2018,Bilitewski2021,Murugan2021,Liu2021,Bertini2025}.
In cellular automata, the spatial sum of the decorrelator is equivalent to the Hamming distance~\cite{Hamming1950}, which serves as the closest analogue of the Lyapunov exponent in discrete-state systems. In Kauffman probabilistic cellular automata, the Hamming distance was instrumental in demonstrating a phase transition between frozen and dynamical regimes~\cite{Derrida86,Weisbuch87}. Moreover, it has also been applied to systematically classify Wolfram rules~\cite{Wolfram83,Alfaro2024}.

Cellular automata provide a convenient playground of minimal models for studying many-body physics. In particular, reversible cellular automata (where each many-body configuration has a unique predecessor) can be viewed as simplified models of Hamiltonian or unitary dynamics. These models have been shown to support generic physical behaviors that interpolate between chaos and integrability.
In one spatial dimension, they have enabled analytical studies of transport~\cite{Buca_2021,Medenjak17,Prosen_2016,Klobas_2022,Klobas21,Wilkinson20,Pozsgay,Prosen2023,Gombor2024,Rustem}, and they have played a crucial role in advancing our understanding of anomalous transport in integrable systems~\cite{Krajnik22,Krajnik24,Krajnik25}. Extending one-dimensional cellular automata to the quantum regime has also led to important results on dynamics, operator growth, and entanglement growth in quantum systems~\cite{Gopalakrishnan2018,Gopalakrishnan2018_2,Alba2019,Klobas_2024,De_Fazio_2024,Klobas2024_2}.
More recently, a new class of reversible cellular automata in higher dimensions, called Momentum-Conserving Parity Check Automata (MCPCA), has been introduced. Remarkably, these systems exhibit a multifractal dynamical structure factor~\cite{Kasim2025}, a phenomenon not previously observed in reversible cellular automata.

In the present work, we explore a new facet of the MCPCA. We show that this model, defined on the square lattice, undergoes a dynamical localization--delocalization phase transition. A crucial feature of the model, which we elaborate on in detail, is the existence of a higher-form (1-form) symmetry: a kind of staggered magnetization around any closed loop is preserved in time. This leads to a natural foliation of the phase space of cell configurations (analog of Hilbert space fragmentation~\cite{Pollmann,Moudgalya_2022}) with respect to the density of the smallest local loop charges. It also imposes important constraints on the dynamics in a manner analogous to local integrals of motion in many-body localization~\cite{Serbyn2013,Abanin19,Sierant_2025}.
We analyze the Hamming distance and the decorrelator to quantify information spreading. These observables clearly distinguish two phases: (i) a localized phase, where local perturbations remain confined, with the decorrelator decaying exponentially in space; and (ii) an extended, chaotic phase, where information spreads ballistically and reaches the system boundaries. Our numerical analysis further indicates that the transition between these phases is of second order, and we provide estimates of the associated critical exponents.

We also address recently observed multifractal dynamical structure factor in MCPCA~\cite{Kasim2025}. It is shown that multifractal scaling of dynamical 2-point function exists on both sides of transition, hence it cannot serve as an order parameter. Finally, we connect the multifractal distribution of the spectral weights to effective (approximate) periodicities of local observables in typical many-body trajectories.

\section{Momentum-conserving parity check cellular automata}
\label{sec:MCPCA}
In this section, we provide a brief overview of momentum-conserving parity check cellular automata (MCPCA), recently introduced in~\cite{Kasim2025}. 
Consider an undirected bipartite graph $G = (E,V)$, where the vertex set $V$ is partitioned into two disjoint subsets, $V = A \cup B$ with $A\cap B=\emptyset$. The edges $e \in E$ connect vertices in $A$ to vertices in $B$. Each edge $e$ carries a dynamical degree of freedom $s_e \in \mathbb{Z}_2$, so that a global state of the automaton is given by $\underline{s} \in \mathbb{Z}_2^{N_{\rm E}}$, with $N_{\rm E} = |E|$. The full phase space is therefore $\mathbb{Z}_2^{N_{\rm E}}$. 

For each vertex $v \in V$, we define a local bijective map $\Phi_v:\mathbb{Z}_2^{n} \to \mathbb{Z}_2^{n}$, acting only on the $n$ edges incident on $v$, denoted $s_1, s_2, \dots, s_n$. To enforce the parity-check condition, the local map satisfies:
\begin{equation}
(s'_1\dots s'_n)=\Phi_v(s_1\dots s_n) 
\; \Rightarrow \;
s_i+s_j=s_i'+s_j'\!\!\pmod{2},
\label{eq:rule1}
\end{equation}
The full update on the graph can then be written as
\begin{equation}\label{full-map}
\Phi = \prod_{v\in B}\Phi_v\prod_{v\in A}\Phi_v,\quad
\underline{s}(t+1) = \Phi (\underline{s}(t)),
\end{equation}
implemented in two parallel steps corresponding to mutually commuting local maps on sublattices $A$ and $B$ \footnote{The local map $\Phi_v$ can as well be identified with a $2^4\times 2^4$ (permutation) matrix which is embedded in $2^{N_{\rm E}}$ dimensional Hilbert space $\mathbb C^{2^{N_{\rm E}}}$ of $N_{\rm E}$ qubits. Similarly, the complete dynamical map $\Phi$ is a $2^{N_{\rm E}} \times 2^{N_{\rm E}}$ matrix representing an element in $S(2^{N_{\rm E}})$, i.e. deterministic dynamics. Let us define observables as functions over phase space $a : \mathbb Z_2^{N_{\rm E}} \to \mathbb C$, or equivalently as vectors in $\mathbb C^{2^{N_{\rm E}}}$. The observable $q$ is a conserved quantity if it satisfies 
$q\circ\Phi = q$, or $q \Phi = q$ in vector notation.}.

In the following, we will focus on the specific regular lattice --- namely the square lattice --- so it is useful to embed the vertices/edges of the graph into $\mathbb{R}^2$. We provide a schematic example of the model on a $4\times 4$ lattice with periodic boundary conditions highlighted in blue (along the $x$-axis) and magenta (along the $y$-axis): 
\begin{equation}
\begin{tikzpicture}[rotate=45]
    \node (n0) at (1,0) {};
    \node (n1) at (1,4.5) {};
    \node (n2) at (2,0) {};
    \node (n3) at (2,4.5) {};
    \node (n4) at (3,0) {};
    \node (n5) at (3,4.5) {};
    \node (n6) at (4,0) {};
    \node (n7) at (4,4.5) {};
    \draw[-] (n0) -- (n1);
    \draw[-] (n2) -- (n3);
    \draw[-] (n4) -- (n5);
    \draw[-] (n6) -- (n7);
    \node (m0) at (0,1) {};
    \node (m1) at (4.5,1) {};
    \node (m2) at (0,2) {};
    \node (m3) at (4.5,2) {};
    \node (m4) at (0,3) {};
    \node (m5) at (4.5,3) {};
    \node (m6) at (0,4) {};
    \node (m7) at (4.5,4) {};
    \draw[-] (m0) -- (m1);
    \draw[-] (m2) -- (m3);
    \draw[-] (m4) -- (m5);
    \draw[-] (m6) -- (m7);
    \node (A1) at (1,1)[circle,draw,fill=black,inner sep=1.5pt]{};
    \node (A2) at (3,1)[circle,draw,fill=black,inner sep=1.5pt]{};
    \node (A3) at (2,2)[circle,draw,fill=black,inner sep=1.5pt]{};
    \node (A4) at (4,2)[circle,draw,fill=black,inner sep=1.5pt]{};
    \node (A5) at (1,3)[circle,draw,fill=black,inner sep=1.5pt]{};
    \node (A6) at (3,3)[circle,draw,fill=black,inner sep=1.5pt]{};
    \node (A7) at (2,4)[circle,draw,fill=black,inner sep=1.5pt]{};
    \node (A8) at (4,4)[circle,draw,fill=black,inner sep=1.5pt]{};
    \node (B1) at (2,1)[circle,draw,fill=blue,inner sep=1.5pt]{};
    \node (B2) at (4,1)[circle,draw,fill=blue,inner sep=1.5pt]{};
    \node (B3) at (1,2)[circle,draw,fill=blue,inner sep=1.5pt]{};
    \node (B4) at (3,2)[circle,draw,fill=blue,inner sep=1.5pt]{};
    \node (B5) at (2,3)[circle,draw,fill=blue,inner sep=1.5pt]{};
    \node (B6) at (4,3)[circle,draw,fill=blue,inner sep=1.5pt]{};
    \node (B7) at (1,4)[circle,draw,fill=blue,inner sep=1.5pt]{};
    \node (B8) at (3,4)[circle,draw,fill=blue,inner sep=1.5pt]{};
    \node (AA) at (5.1,3.5)[circle,draw,fill=black,inner sep=1.5pt]{};
    \node (AAt) at (5.95,3.5) {$v \in A$};
    \node (BB) at (5.1,2.5)[circle,draw,fill=blue,inner sep=1.5pt]{};
    \node (BBt) at (6.05,2.5) {$v \in B$};
    \node (CC1) at (5.3,1.5)[circle,draw,fill=white,inner sep=2.5pt]{};
    \node (CC2) at (5.75,1.05)[circle,draw,fill=red!50,inner sep=2.5pt]{};
    \node (CCt) at (5.85,1.55) {$s_e$};
    \node (C1) at (1.5,1)[circle,draw,fill=red!50,inner sep=2.5pt]{};
    \node (C2) at (2.5,1)[circle,draw,fill=red!50,inner sep=2.5pt]{};
    \node (C3) at (3.5,1)[circle,draw,fill=white,inner sep=2.5pt]{};
    \node (C5) at (1,1.5)[circle,draw,fill=white,inner sep=2.5pt]{};
    \node (C6) at (2,1.5)[circle,draw,fill=white,inner sep=2.5pt]{};
    \node (C7) at (3,1.5)[circle,draw,fill=red!50,inner sep=2.5pt]{};
    \node (C8) at (4,1.5)[circle,draw,fill=red!50,inner sep=2.5pt]{};
    \node (D1) at (1.5,2)[circle,draw,fill=red!50,inner sep=2.5pt]{};
    \node (D2) at (2.5,2)[circle,draw,fill=white,inner sep=2.5pt]{};
    \node (D3) at (3.5,2)[circle,draw,fill=white,inner sep=2.5pt]{};
    \node (D5) at (1,2.5)[circle,draw,fill=red!50,inner sep=2.5pt]{};
    \node (D6) at (2,2.5)[circle,draw,fill=red!50,inner sep=2.5pt]{};
    \node (D7) at (3,2.5)[circle,draw,fill=white,inner sep=2.5pt]{};
    \node (D8) at (4,2.5)[circle,draw,fill=white,inner sep=2.5pt]{};
    \node (E1) at (1.5,3)[circle,draw,fill=white,inner sep=2.5pt]{};
    \node (E2) at (2.5,3)[circle,draw,fill=red!50,inner sep=2.5pt]{};
    \node (E3) at (3.5,3)[circle,draw,fill=red!50,inner sep=2.5pt]{};
    \node (E5) at (1,3.5)[circle,draw,fill=red!50,inner sep=2.5pt]{};
    \node (E6) at (2,3.5)[circle,draw,fill=white,inner sep=2.5pt]{};
    \node (E7) at (3,3.5)[circle,draw,fill=red!50,inner sep=2.5pt]{};
    \node (E8) at (4,3.5)[circle,draw,fill=red!50,inner sep=2.5pt]{};
    \node (F1) at (1.5,4)[circle,draw,fill=white,inner sep=2.5pt]{};
    \node (F2) at (2.5,4)[circle,draw,fill=white,inner sep=2.5pt]{};
    \node (F3) at (3.5,4)[circle,draw,fill=red!50,inner sep=2.5pt]{};
    \draw (1,4.3) arc (180:0:0.1cm);
    \draw (2,4.3) arc (180:0:0.1cm);
    \draw (3,4.3) arc (180:0:0.1cm);
    \draw (4,4.3) arc (180:0:0.1cm);
    \draw (1.2,0.2) arc (0:-180:0.1cm);
    \draw (2.2,0.2) arc (0:-180:0.1cm);
    \draw (3.2,0.2) arc (0:-180:0.1cm);
    \draw (4.2,0.2) arc (0:-180:0.1cm);
    \draw (4.3,1) arc (90:-90:0.1cm);
    \draw (4.3,2) arc (90:-90:0.1cm);
    \draw (4.3,3) arc (90:-90:0.1cm);
    \draw (4.3,4) arc (90:-90:0.1cm);
    \draw (0.2,1) arc (90:270:0.1cm);
    \draw (0.2,2) arc (90:270:0.1cm);
    \draw (0.2,3) arc (90:270:0.1cm);
    \draw (0.2,4) arc (90:270:0.1cm);
    \node (G1) at (0.5,1)[circle,draw,fill=red!50,inner sep=2.5pt]{};
    \node (G2) at (0.5,2)[circle,draw,fill=white,inner sep=2.5pt]{};
    \node (G3) at (0.5,3)[circle,draw,fill=red!50,inner sep=2.5pt]{};
    \node (G4) at (0.5,4)[circle,draw,fill=white,inner sep=2.5pt]{};
    \node (G5) at (1,0.5)[circle,draw,fill=red!50,inner sep=2.5pt]{};
    \node (G6) at (2,0.5)[circle,draw,fill=white,inner sep=2.5pt]{};
    \node (G7) at (3,0.5)[circle,draw,fill=red!50,inner sep=2.5pt]{};
    \node (G8) at (4,0.5)[circle,draw,fill=white,inner sep=2.5pt]{};
    \draw[thick,line width=1.3mm,opacity=0.35,color=violet] (1.2,0.2) arc[start angle=360, end angle=180, radius=0.1]  -- (1,4.3) arc[start angle=180, end angle=0, radius=0.1];
    \draw[thick,line width=1.3mm,opacity=0.35,color=blue] (0.2,0.8) arc[start angle=-90, end angle=-270, radius=0.1] -- (4.3,1) arc[start angle=90, end angle=-90, radius=0.1];
\end{tikzpicture}
\label{eq:square_lattice}
\end{equation}

For the square lattice, the most symmetric parity check conserving reversible automaton that will be considered here, is defined as: $\Phi_v(s,s,s,s) = (s,s,s,s)$ and otherwise
$\Phi_v(s,s',s'',s''') = (\bar{s},\bar{s}',\bar{s}'',\bar{s}''')$, where $\bar{s}=1-s$, illustrated graphically as:
\begin{eqnarray}
	\begin{tikzpicture}[baseline={(current bounding box.center)},every node/.style={inner sep=0,outer sep=0},line cap=rect,scale=0.5]
    \node (n0) at (0,0)[circle,draw,fill,inner sep=1.25pt] {};
    \node (n1) at (1,1) {};
    \node (n2) at (-1,-1) {};
    \node (n3) at (1,-1) {};
    \node (n4) at (-1,1) {};
    \draw[-] (n0) -- (n1);
    \draw[-] (n0) -- (n2);
    \draw[-] (n0) -- (n3);
    \draw[-] (n0) -- (n4);
    \node (A) at (0.5,0.5)[circle,draw,fill=white,inner sep=2.5pt]{};
    \node (B) at (0.5,-0.5)[circle,,draw,fill=white,inner sep=2.5pt]{};
    \node (C) at (-0.5,-0.5)[circle,draw,fill=white,inner sep=2.5pt]{};
    \node (D) at (-0.5,0.5)[circle,,draw,fill=white,inner sep=2.5pt]{};
    \end{tikzpicture}
	\leftrightarrow
	\begin{tikzpicture}[baseline={(current bounding box.center)},every node/.style={inner sep=0,outer sep=0},line cap=rect,scale=0.5]
    \node (n0) at (0,0)[circle,draw,fill,inner sep=1.25pt] {};
    \node (n1) at (1,1) {};
    \node (n2) at (-1,-1) {};
    \node (n3) at (1,-1) {};
    \node (n4) at (-1,1) {};
    \draw[-] (n0) -- (n1);
    \draw[-] (n0) -- (n2);
    \draw[-] (n0) -- (n3);
    \draw[-] (n0) -- (n4);
    \node (A) at (0.5,0.5)[circle,draw,fill=white,inner sep=2.5pt]{};
    \node (B) at (0.5,-0.5)[circle,,draw,fill=white,inner sep=2.5pt]{};
    \node (C) at (-0.5,-0.5)[circle,draw,fill=white,inner sep=2.5pt]{};
    \node (D) at (-0.5,0.5)[circle,,draw,fill=white,inner sep=2.5pt]{};
    \end{tikzpicture},
    &\quad&
	\begin{tikzpicture}[baseline={(current bounding box.center)},every node/.style={inner sep=0,outer sep=0},line cap=rect,scale=0.5]
    \node (n0) at (0,0)[circle,draw,fill,inner sep=1.25pt] {};
    \node (n1) at (1,1) {};
    \node (n2) at (-1,-1) {};
    \node (n3) at (1,-1) {};
    \node (n4) at (-1,1) {};
    \draw[-] (n0) -- (n1);
    \draw[-] (n0) -- (n2);
    \draw[-] (n0) -- (n3);
    \draw[-] (n0) -- (n4);
    \node (A) at (0.5,0.5)[circle,draw,fill=red!50,inner sep=2.5pt]{};
    \node (B) at (0.5,-0.5)[circle,,draw,fill=red!50,inner sep=2.5pt]{};
    \node (C) at (-0.5,-0.5)[circle,draw,fill=red!50,inner sep=2.5pt]{};
    \node (D) at (-0.5,0.5)[circle,,draw,fill=red!50,inner sep=2.5pt]{};
    \end{tikzpicture}
	\leftrightarrow
	\begin{tikzpicture}[baseline={(current bounding box.center)},every node/.style={inner sep=0,outer sep=0},line cap=rect,scale=0.5]
    \node (n0) at (0,0)[circle,draw,fill,inner sep=1.25pt] {};
    \node (n1) at (1,1) {};
    \node (n2) at (-1,-1) {};
    \node (n3) at (1,-1) {};
    \node (n4) at (-1,1) {};
    \draw[-] (n0) -- (n1);
    \draw[-] (n0) -- (n2);
    \draw[-] (n0) -- (n3);
    \draw[-] (n0) -- (n4);
    \node (A) at (0.5,0.5)[circle,draw,fill=red!50,inner sep=2.5pt]{};
    \node (B) at (0.5,-0.5)[circle,,draw,fill=red!50,inner sep=2.5pt]{};
    \node (C) at (-0.5,-0.5)[circle,draw,fill=red!50,inner sep=2.5pt]{};
    \node (D) at (-0.5,0.5)[circle,,draw,fill=red!50,inner sep=2.5pt]{};
    \end{tikzpicture},
    \quad
    \begin{tikzpicture}[baseline={(current bounding box.center)},every node/.style={inner sep=0,outer sep=0},line cap=rect,scale=0.5]
    \node (n0) at (0,0)[circle,draw,fill,inner sep=1.25pt] {};
    \node (n1) at (1,1) {};
    \node (n2) at (-1,-1) {};
    \node (n3) at (1,-1) {};
    \node (n4) at (-1,1) {};
    \draw[-] (n0) -- (n1);
    \draw[-] (n0) -- (n2);
    \draw[-] (n0) -- (n3);
    \draw[-] (n0) -- (n4);
    \node (A) at (0.5,0.5)[circle,draw,fill=red!50,inner sep=2.5pt]{};
    \node (B) at (0.5,-0.5)[circle,,draw,fill=white,inner sep=2.5pt]{};
    \node (C) at (-0.5,-0.5)[circle,draw,fill=white,inner sep=2.5pt]{};
    \node (D) at (-0.5,0.5)[circle,,draw,fill=white,inner sep=2.5pt]{};
    \end{tikzpicture}
	\leftrightarrow
	\begin{tikzpicture}[baseline={(current bounding box.center)},every node/.style={inner sep=0,outer sep=0},line cap=rect,scale=0.5]
    \node (n0) at (0,0)[circle,draw,fill,inner sep=1.25pt] {};
    \node (n1) at (1,1) {};
    \node (n2) at (-1,-1) {};
    \node (n3) at (1,-1) {};
    \node (n4) at (-1,1) {};
    \draw[-] (n0) -- (n1);
    \draw[-] (n0) -- (n2);
    \draw[-] (n0) -- (n3);
    \draw[-] (n0) -- (n4);
    \node (A) at (0.5,0.5)[circle,draw,fill=white,inner sep=2.5pt]{};
    \node (B) at (0.5,-0.5)[circle,,draw,fill=red!50,inner sep=2.5pt]{};
    \node (C) at (-0.5,-0.5)[circle,draw,fill=red!50,inner sep=2.5pt]{};
    \node (D) at (-0.5,0.5)[circle,,draw,fill=red!50,inner sep=2.5pt]{};
    \end{tikzpicture}
    \nonumber \\
    \begin{tikzpicture}[baseline={(current bounding box.center)},every node/.style={inner sep=0,outer sep=0},line cap=rect,scale=0.5]
    \node (n0) at (0,0)[circle,draw,fill,inner sep=1.25pt] {};
    \node (n1) at (1,1) {};
    \node (n2) at (-1,-1) {};
    \node (n3) at (1,-1) {};
    \node (n4) at (-1,1) {};
    \draw[-] (n0) -- (n1);
    \draw[-] (n0) -- (n2);
    \draw[-] (n0) -- (n3);
    \draw[-] (n0) -- (n4);
    \node (A) at (0.5,0.5)[circle,draw,fill=white,inner sep=2.5pt]{};
    \node (B) at (0.5,-0.5)[circle,,draw,fill=red!50,inner sep=2.5pt]{};
    \node (C) at (-0.5,-0.5)[circle,draw,fill=white,inner sep=2.5pt]{};
    \node (D) at (-0.5,0.5)[circle,,draw,fill=red!50,inner sep=2.5pt]{};
    \end{tikzpicture}
	\leftrightarrow
	\begin{tikzpicture}[baseline={(current bounding box.center)},every node/.style={inner sep=0,outer sep=0},line cap=rect,scale=0.5]
    \node (n0) at (0,0)[circle,draw,fill,inner sep=1.25pt] {};
    \node (n1) at (1,1) {};
    \node (n2) at (-1,-1) {};
    \node (n3) at (1,-1) {};
    \node (n4) at (-1,1) {};
    \draw[-] (n0) -- (n1);
    \draw[-] (n0) -- (n2);
    \draw[-] (n0) -- (n3);
    \draw[-] (n0) -- (n4);
    \node (A) at (0.5,0.5)[circle,draw,fill=red!50,inner sep=2.5pt]{};
    \node (B) at (0.5,-0.5)[circle,,draw,fill=white,inner sep=2.5pt]{};
    \node (C) at (-0.5,-0.5)[circle,draw,fill=red!50,inner sep=2.5pt]{};
    \node (D) at (-0.5,0.5)[circle,,draw,fill=white,inner sep=2.5pt]{};
    \end{tikzpicture},
    &\quad&
	\begin{tikzpicture}[baseline={(current bounding box.center)},every node/.style={inner sep=0,outer sep=0},line cap=rect,scale=0.5]
    \node (n0) at (0,0)[circle,draw,fill,inner sep=1.25pt] {};
    \node (n1) at (1,1) {};
    \node (n2) at (-1,-1) {};
    \node (n3) at (1,-1) {};
    \node (n4) at (-1,1) {};
    \draw[-] (n0) -- (n1);
    \draw[-] (n0) -- (n2);
    \draw[-] (n0) -- (n3);
    \draw[-] (n0) -- (n4);
    \node (A) at (0.5,0.5)[circle,draw,fill=white,inner sep=2.5pt]{};
    \node (B) at (0.5,-0.5)[circle,,draw,fill=white,inner sep=2.5pt]{};
    \node (C) at (-0.5,-0.5)[circle,draw,fill=red!50,inner sep=2.5pt]{};
    \node (D) at (-0.5,0.5)[circle,,draw,fill=red!50,inner sep=2.5pt]{};
    \end{tikzpicture}
	\leftrightarrow
	\begin{tikzpicture}[baseline={(current bounding box.center)},every node/.style={inner sep=0,outer sep=0},line cap=rect,scale=0.5]
    \node (n0) at (0,0)[circle,draw,fill,inner sep=1.25pt] {};
    \node (n1) at (1,1) {};
    \node (n2) at (-1,-1) {};
    \node (n3) at (1,-1) {};
    \node (n4) at (-1,1) {};
    \draw[-] (n0) -- (n1);
    \draw[-] (n0) -- (n2);
    \draw[-] (n0) -- (n3);
    \draw[-] (n0) -- (n4);
    \node (A) at (0.5,0.5)[circle,draw,fill=red!50,inner sep=2.5pt]{};
    \node (B) at (0.5,-0.5)[circle,,draw,fill=red!50,inner sep=2.5pt]{};
    \node (C) at (-0.5,-0.5)[circle,draw,fill=white,inner sep=2.5pt]{};
    \node (D) at (-0.5,0.5)[circle,,draw,fill=white,inner sep=2.5pt]{};
    \end{tikzpicture},
    \quad
	+ \frac{\pi}{2} \: \mathrm{rotations}
    \label{rules_square}
\end{eqnarray}
The rule above has an additional property of momentum conservation.
Viewing the excitation on the edge $e$ as a unit mass and unit velocity particle carrying directed momentum $\vec{e}$ (using the convention that the particle is moving from sublattice $A$ to sublattice $B$), one finds:
\begin{equation}
\label{eq:momentum_conservation}
    (s'_1\dots s'_n)=\Phi_v(s_1\dots s_n), \quad \sum_{e=1}^n s_e \vec{e} = -\sum_{e=1}^n s'_e \vec{e} 
\end{equation}
The two conditions (\ref{eq:rule1},\ref{eq:momentum_conservation}) give the general definition of 
momentum conserving partiy check automaton (MCPCA),
while (\ref{rules_square}) considered here is the most symmetric example.
 
Let us define local basis-observables $Z_e$ as $Z_e(\underline{s}) = (-1)^{s_e}$.
The crucial feature of our local dynamical map $\Phi_{v}$ is the fact that for any pair of degrees of freedom $s_e$ and $s_{e'}$, on which $\Phi_{v}$ acts, it flips the sign of the associated staggered magnetization:
\begin{equation}
 (Z_e-Z_{e'})\circ \Phi_v = -Z_e + Z_{e'},\quad \forall e,e' \in\{1,\ldots,n\}.
\end{equation}
Using this fundamental relation it is possible to construct the set of local conserved quantities for the full dynamical map~(\ref{full-map}). To any \textit{closed} loop $\gamma = (e_1 , e_2 , ... , e_{|\gamma|})$ on the graph $G$, of even length $|\gamma|$ , one can associate a charge: 
\begin{equation}\label{loop-charge}
    M_{\gamma} =\frac{1}{2} \sum_{j=1}^{|\gamma|} (-1)^{j-1} Z_{e_j}
\end{equation}
that is fixed by the dynamical map 
\begin{equation}
M_{\gamma} \circ \Phi = M_\gamma,
\end{equation} i.e. the value of this charge is conserved in time for any closed loop $\gamma$. In the following, we will refer to $M_\gamma$ as the loop charges. Actually, this automaton belongs to a more general class of systems that possess an extensive set of local conserved quantities --- 1-form symmetries --- which can be constructed as staggered-magnetization operators along all closed loops of the graph. This class of systems is introduced and discussed in more details in related work of our group \cite{PavelOrlov}.

As a direct consequence of the loop charges, the following three properties can be derived. Firstly, the parity $\pi_{\gamma} = (-1)^{\sum_{k=1}^{|\gamma|} s_{e_k} }$ of a sum of dynamical variables along a loop $\gamma$ can be expressed as $\pi_{\gamma} = e^{i \pi M_k}$, and as a consequence it is also conserved. Secondly, the two $Néel$ configurations on a loop $(s_{e_1} , ... , s_{e_{|\gamma|} } ) = (0,1,0,1, ... , 0,1 )$ and $(1,0,1,0,...,1,0)$ correspond to the highest $+|\gamma|$ and the lowest $-|\gamma|$ values of the charge $M_{\gamma}$. As a result, such configurations are frozen under the dynamics. And finally, the staggered magnetization around the lattice (with periodic boundary conditions assumed) corresponds to the momentum defined in~\eqref{eq:momentum_conservation}, for example on the square lattice of size $N\times N$, we can write the conserved $x$-momentum at a constant value of any y-coordinate $Y$ as $\frac{1}{2}\sum_{x=1}^N (-1)^{x-1}Z_{x,Y}$, and similarly for the $y-$momenta. 

These three properties of the dynamical map $\Phi$ were previously identified in~\cite{Kasim2025}. Here, we emphasize that they all follow from the conservation of the loop charges defined in~\eqref{loop-charge}, i.e. from 1-form symmetry. However, our primary aim is to investigate how these conserved quantities affect the dynamical behavior of the system.

It is clear that even if we can define a loop-charge~\eqref{loop-charge} for any $\gamma$, not all loop charges are independent of each other. It is then important to fix a basis of charges such that for any $\gamma$ the charge $M_{\gamma}$ could be expressed as a linear combination of these basis charges. For example, on the square lattice with periodic boundary conditions, one can pick loops around all the plaquettes of the lattice: 

\begin{eqnarray}
\begin{tikzpicture}[baseline={(current bounding box.center)},every node/.style={inner sep=0,outer sep=0},line cap=rect,scale=1.25]
	\node (n0) at (0,0)[circle,fill,inner sep=1.25pt] {};
    \node (n1) at (-0.375,-0.375) {};
    \node (n2) at (0.75,0.75)[circle,draw,fill=blue,inner sep=1.25pt] {};
	\node (n3) at (-0.75,0.75)[circle,draw,fill=blue,inner sep=1.25pt] {};
    \node (n4) at (0,1.5)[circle,fill,inner sep=1.25pt] {};
	\node (n5) at (0.375,-0.375) {};
	\node (n6) at (-1.125,0.375) {};
	\node (n7) at (-1.125,1.125) {};
	\node (n8) at (1.125,0.375) {};
	\node (n9) at (1.125,1.125) {};
	\node (n10) at (-0.375,1.875) {};
	\node (n11) at (0.375,1.875) {};
    \draw[-] (n0) -- (n1);
    \draw[-] (n0) -- (n2);
    \draw[-] (n0) -- (n3);
    \draw[-] (n0) -- (n5);
    \draw[-] (n3) -- (n4);
    \draw[-] (n2) -- (n4);
    \draw[-] (n3) -- (n6);
    \draw[-] (n3) -- (n7);
    \draw[-] (n2) -- (n8);
    \draw[-] (n2) -- (n9);
    \draw[-] (n4) -- (n10);
    \draw[-] (n4) -- (n11);
    \node[scale=2] (ar) at (0,0.75) {$\circlearrowleft$};
	\draw[black,fill=red!50] (0.375cm+3.535533906pt,0.375cm+3.535533906pt) arc  (45:225:5pt); 
	\draw[black,fill=white] (0.375cm-3.535533906pt,0.375cm-3.535533906pt) arc  (225:405:5pt);
	\draw[black,fill=red!50] (-0.375cm+3.535533906pt,0.375cm+3.535533906pt) arc  (45:225:5pt); 
	\draw[black,fill=white] (-0.375cm-3.535533906pt,0.375cm-3.535533906pt) arc  (225:405:5pt);
    \draw[black,fill=red!50] (0.375cm+3.535533906pt,1.125cm+3.535533906pt) arc  (45:225:5pt); 
	\draw[black,fill=white] (0.375cm-3.535533906pt,1.125cm-3.535533906pt) arc  (225:405:5pt);
    \draw[black,fill=red!50] (-0.375cm+3.535533906pt,1.125cm+3.535533906pt) arc  (45:225:5pt); 
	\draw[black,fill=white] (-0.375cm-3.535533906pt,1.125cm-3.535533906pt) arc  (225:405:5pt);
	\end{tikzpicture}
	\label{eq:parity}
\end{eqnarray}
in addition to any of the 2 {\emph{topological}} loops that wrap around the $x$ and $y$ directions (see the highlighted parts in~\eqref{eq:square_lattice}).

Such square plaquette has 5 possible values of $M_\gamma \in  \{-2,-1,0,1,2\}$. On a lattice of size $N\times N$ there are $N_{\rm E}=2N^2$ particles and $N^2$ conserved plaquette loop charges. The dynamics of the initial state will depend greatly on the distribution of the values of the staggered magnetization over the plaquettes in this state. This comes from the correspondence of the size of the configuration space with the values of the conserved quantities. If all the plaquettes are set to $M_\gamma=\pm2$ (the two Néel states: $(0,1,0,1),(1,0,1,0)$), such a configuration will have no dynamics and one possible (frozen) state. On the other hand, setting all the plaquettes to $M_\gamma=0$ will correspond to the largest size of configuration space, since for $M_\gamma=0$ a plaquette has the highest number of configurations (6 out of 16). 

Since the square-lattice plaquette charges depend only on the $4$ bits, it is not possible to define their densities as a set of macroscopic degrees of freedom\footnote{In contrast with topological loop charges, which act on $N$ degrees of freedom, and so, the density $M_{\gamma}/N$ for these charges can be well-defined in the thermodynamic limit}. Instead, we define other macroscopic quantities that would incorporate some global information about the distribution of plaquette charges in a state. The simplest example of such quantities are the plaquette densities:
\begin{equation}
    n_q = \frac{1}{N^2} (\#\text{plaquettes} \; \text{s.t.} \; M_{\gamma} = q  ), \quad q \in \{-2,\ldots,2\}.
\end{equation}
Note that such densities are also conserved in time and, obviously, sum up to unity $\sum_{q} n_q =1$. Working with ensembles of states that have particular values of these densities, we will study how their values drive the dynamics of the system. We can expect that as the values of $n_{\pm2}$ and $n_{\pm1}$ increase, the dynamics will become slower and slower. But to characterize this quantitatively, we will utilize two information-theoretical quantities: the Hamming distance and the decorrelator. In the next section, we discuss their definitions and basic properties.

\section{Sampling of the density of staggered magnetization}
\label{sec:Stag_mag_den}

We introduced the notion of plaquettes density in sec~\ref{sec:MCPCA}. We want to find a natural way to select, or sample over a specific sector with specific plaquette configuration. 
A state of MCPCA on a square lattice of size $N\times N$ can be encoded in a bitstring of size $2N^2$. In the maximal entropy state, as considered in Ref.~\cite{Kasim2025}, all the bits in the bitstring had equal probabilities of being $0$ or $1$. This corresponds to a time-translation invariant, i.e. equilibrium state of the system with maximal (infinite) effective temperature.

\begin{figure}
    \centering
    \includegraphics[width=0.9\linewidth]{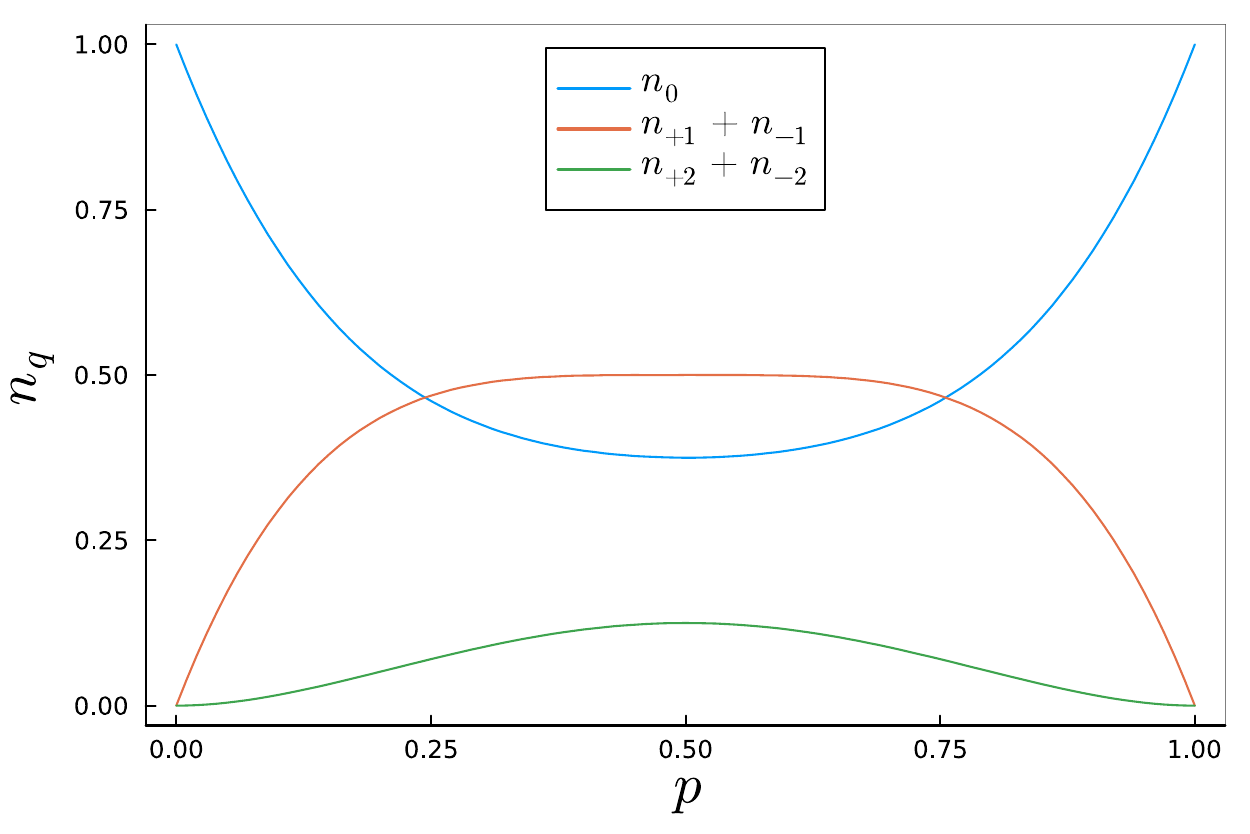}
    \caption{The density of plaquettes $n_q$ as a function of the sampling probability $p$. Note that $n_{+1} = n_{-1}$ and $n_{+2}=n_{-2}$ we thus plot the sums of these densities.}
    \label{fig:sampling_probability}
\end{figure}

 Here, we will instead generate an ensemble of initial states with a general Bernoulli process where each bit i.i.d. variable with probability $p_0=p$ of having value $0$ and probability $p_1=1-p$ of having value $1$. We note that this state is preserved under time-evolution only for $p=1/2$, otherwise it undergoes a non-trival dynamics under MCPCA. In Fig.~\ref{fig:sampling_probability}, we show the densities of plaquettes $n_q$ in the Bernoulli initial states as a function of the probability $p$. We see that depending on the sampling probability $p$, we can to some extent control the average staggered magnetization densities, which in turn, are conserved with time evolution.  

It is useful to note that one can also precisely select the sector of the configuration space with specific values of $\{n_q\}$ by using an adaptation of Metropolis (aka Markov Chain Monte Carlo) algorithm. We give a quick outline for such an algorithm: (i) start by taking a random initial state - cell configuration $\underline{s}$, (ii) repeatedly flip randomly chosen bits while accepting the flips with probability $\max\{1,
\rho(\underline{s})/\rho(\underline{s}_{\rm new})\}$ using the following generalized Gibbs cost function:
\begin{equation}
    \rho(\underline{s}) = Z^{-1}\exp \left(-\sum_q \beta_q N_q[\underline{s}] \right),
\end{equation}
where $N_q[\underline{s}]$ is the number of plaquettes with a loop charge $q$ in the specific configuration $\underline{s}$,
(iii) while this-way simulating a large $M$ Markov chain 
ensemble $\{\underline{s}_k\}_{k=1}^{M}$, monitor the
mean values $\langle n_q\rangle = \frac{1}{N_{\rm E} M}\sum_{k=1}^M N_q[\underline{s}_k]$.
The coefficients $\beta_q$ are the corresponding generalized chemical potentials which can be tuned to target specific $\{n_q\}$, that is, determining $\boldsymbol{\beta}(\boldsymbol{n})$ by empirically inverting $\boldsymbol{n}(\boldsymbol{\beta})$. 
This method should give access to the full $4$D equilibrium phase diagram of the MCPCA for the square lattice. In the current work, we choose to use the
simpler one-parametric Bernoulli model which is rich enough to display the transition between two phases and leave exploration of the full phase diagram for future work.

\section{The Hamming distance and the phase transition}
\label{sec:Hamming_transition}
For two configurations on a $N\times N$ square lattice $ \underline{s}^{A} = \{ s_{\mathbf{r}}^{A} \}$ and $ \underline{s}^{B} = \{s_{\mathbf{r}}^{B} \} $ (here and elsewhere $\mathbf{r} = (x,y)\in \mathbb Z_N^2$ are coordinates on a lattice) the Hamming distance is defined as the number of bits differing between the two configurations:
\begin{equation}\label{eq:hamming}
    H_{d}(\underline{s}^{A},\underline{s}^{B}) \equiv \sum_{\mathbf{r}} |s_{\mathbf{r}}^{A} - s_{\mathbf{r}}^{B}|. 
\end{equation}
Interpreting $\underline{s}^{A}$ and $\underline{s}^{B}$ as two different initial conditions for $t=0$ in our cellular automata, one can study the evolution of the Hamming distance with time 
\begin{equation}
    H_{d} (t)=H_d(\underline{s}^A(t),\underline{s}^B(t))\,,\quad
\underline{s}(t) = \Phi^{(t)}(\underline{s})\,.
\end{equation} 
We will be mostly interested in the situation where $\underline{s}^{A}$ and $\underline{s}^{B}$ differ by one bit, i.e. initially $H_{d}(0) = 1$. In this setting, $H_d(t)$ tracks how a single-bit error spreads through the system, directly analogous to the butterfly effect in classical chaotic dynamics~\cite{Bilitewski2018,Das2018,Bilitewski2021}

In practice, we generate $\underline{s}^{A}$ from a chosen ensemble, specified for instance by the set of densities ${n_q}$ or by a Bernoulli sampling parameter $p$. The perturbed configuration $\underline{s}^{B}$ is then obtained by flipping a single randomly selected bit in $\underline{s}^{A}$. We then average $H_d(t)$ over both the ensemble of initial conditions and the random choice of the flipped bit.

The dynamics of the Hamming distance, $H_d(t)$, generally proceeds in two stages. At early times, it grows monotonically, with the growth rate bounded by $t^{2}$ on a two-dimensional lattice—a consequence of the locality of the dynamical map. At later times, $H_d(t)$ saturates to a steady-state value that can be expressed as the long-time average
\begin{equation}
    \bar{H}_d = \lim_{T \rightarrow \infty }\frac{1}{T} \sum_{t=0}^{T-1}H_d(t),
\end{equation}

\begin{figure}[ ]
    \centering
    \includegraphics[width=0.95\linewidth]{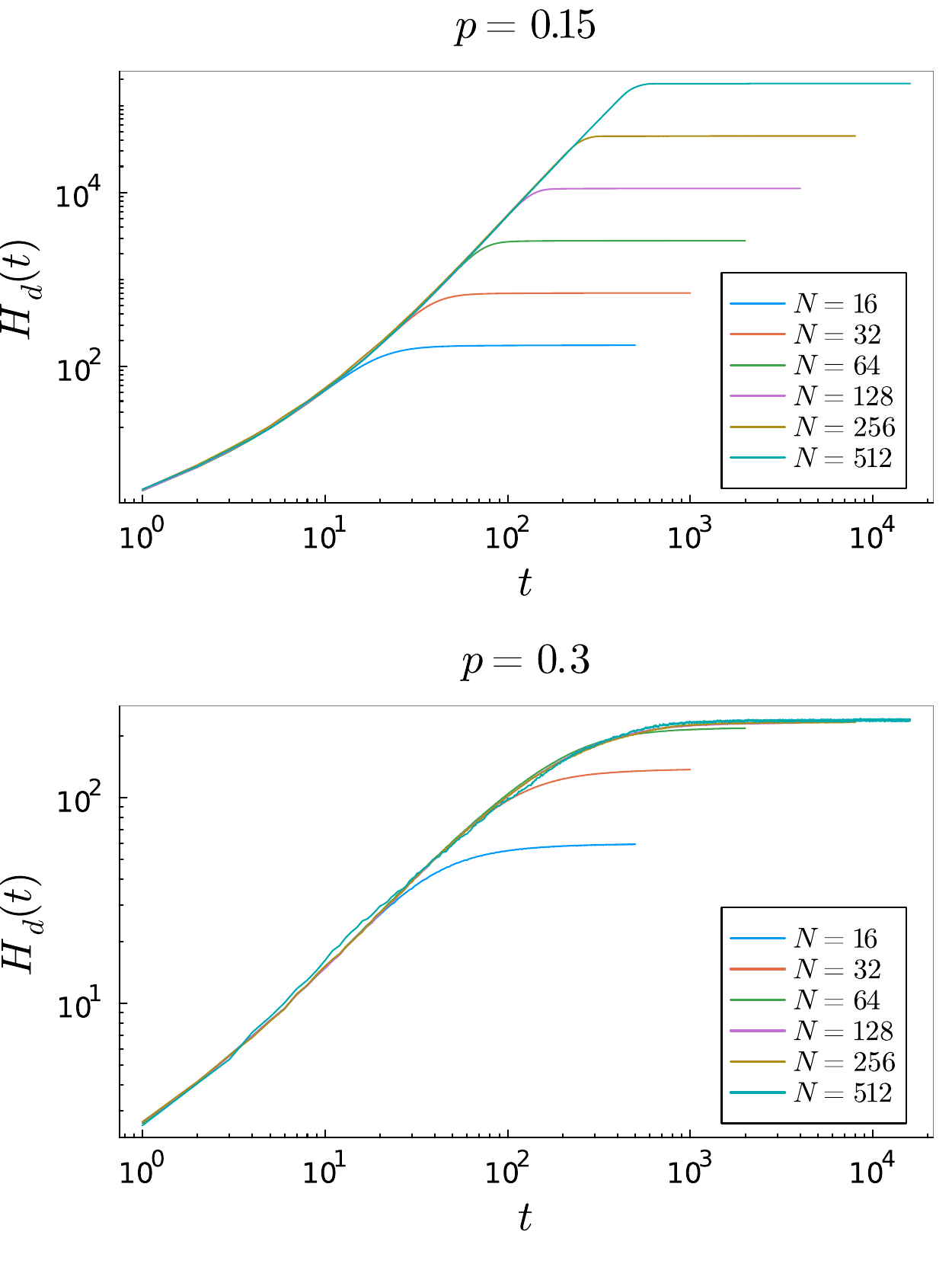}
    \caption{The Hamming distance as a function of time $H_d(t)$ for different lattice sizes $N\times N$ and for $2$ different values of the sampling probability $p$. The top panel is in the delocalized phase (note equal steps in $\log\bar{H}_d$ for a geometric sequence of $N$), while the lower panel is in the localized phase.}
    \label{fig:Hamming_distance_different_NS}
\end{figure}

The saturation value can be extensive in the number of degrees of freedom $ \bar{H}_d \sim N^{2}$. For example, this is the case for the top panel of Fig.~\ref{fig:Hamming_distance_different_NS}, where the initial microscopic error in one cell spreads on a macroscopic scale. If an ensemble of states has this property, we will label this ensemble as belonging to the phase of extended information spreading. 

Conversely, if the saturation value is subextensive $\bar{H}_d \sim N^{\alpha}$, $\alpha<2$, the information about the initial local error remains confined and fails to propagate macroscopically. This behavior is illustrated in the lower panel of Fig.~\ref{fig:Hamming_distance_different_NS}, where $\bar{H}_d$ shows little or no dependence on system size, corresponding to the case $\alpha = 0$. We refer to this regime as the phase of localized information spreading.

Taking the above definitions of these two phases, one can observe that the density of the saturation value of the Hamming distance $h_{d} = \lim_{N \rightarrow \infty} \frac{1}{2N^2} \bar{H}_{d} $ can be interpreted as a convenient {\em order parameter}:
\begin{equation}
\begin{aligned}
    \text{ Delocalized Information Phase} \quad : \quad h_{d}>0  \\
    \text{ Localized Information Phase} \quad : \quad h_{d} = 0
\end{aligned}
\end{equation}
in analogy with the magnetization density in the ferro-to-para-magnet 2nd order phase transition.

As Fig.~\ref{fig:Hamming_distance_different_NS} already illustrates, for small enough values of the sampling parameter $p$, where the plaquette charges with $q=0$ dominate, $n_0\gg n_{\pm 1,\pm 2}$ (see Fig.~\ref{fig:sampling_probability}), our system belongs to the delocalized phase. For large enough values of $p \le 1/2$, where the dominance of $q=\pm1$ and $q=\pm2$ plaquettes takes place, $n_{\pm 1,\pm 2} > n_0$, the system is localized. This indicates a possibility of critical scaling in the transition between the two phases. 

\begin{figure}[ ]
    \centering
    \includegraphics[width=0.95\linewidth]{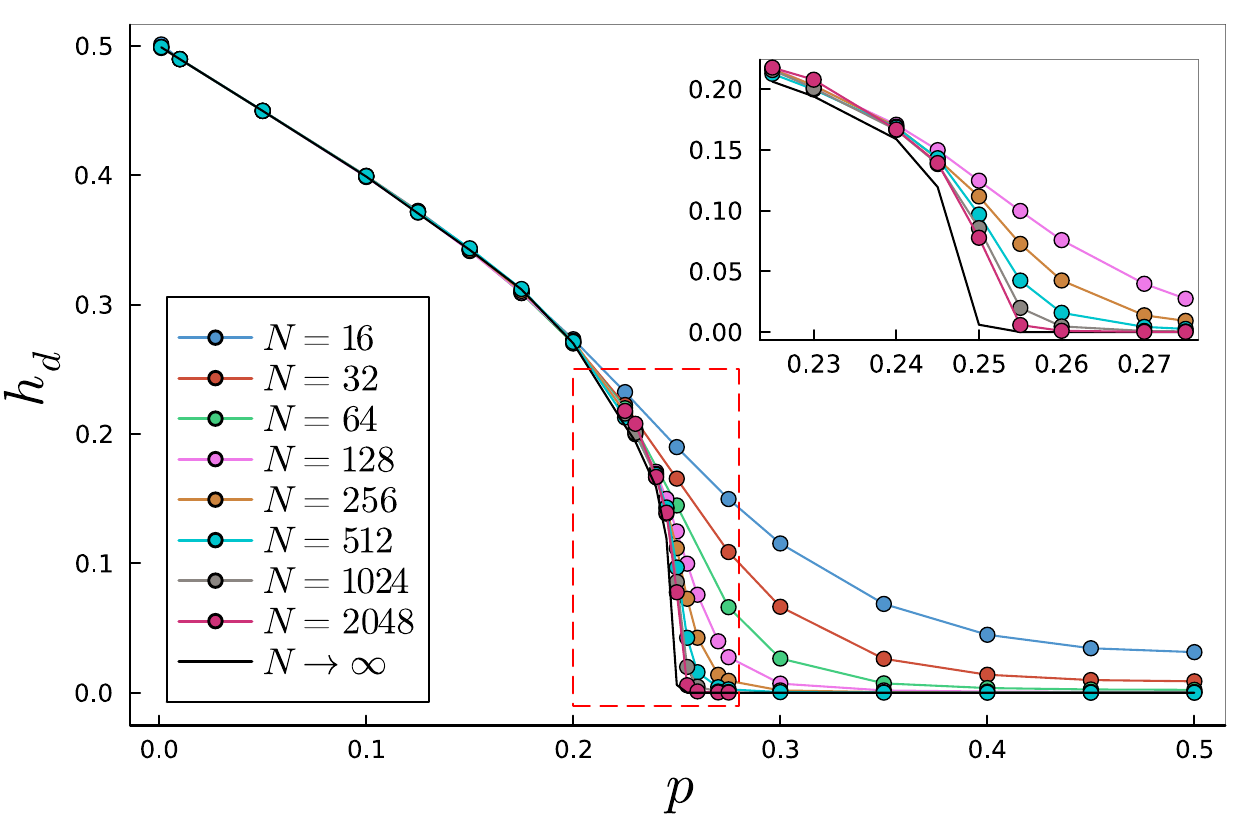}
    \caption{The density of the Hamming distance as a function of the sampling probability $p$ for different lattice sizes $N\times N$. The density of Hamming distance converges to nonzero values for $p<0.25$, while for $p>0.25$ it goes to $0$. The inset is the zoomed data for $p\in [0.225,0.275]$ (the red box). The black line shows an extrapolation for $N\to \infty$ found by linear fitting of $h_d$ as a function of $1/\ln N$.}
    \label{fig:Hamming_transition}
\end{figure}

\begin{figure}[ ]
    \centering
    \includegraphics[width=0.95\linewidth]{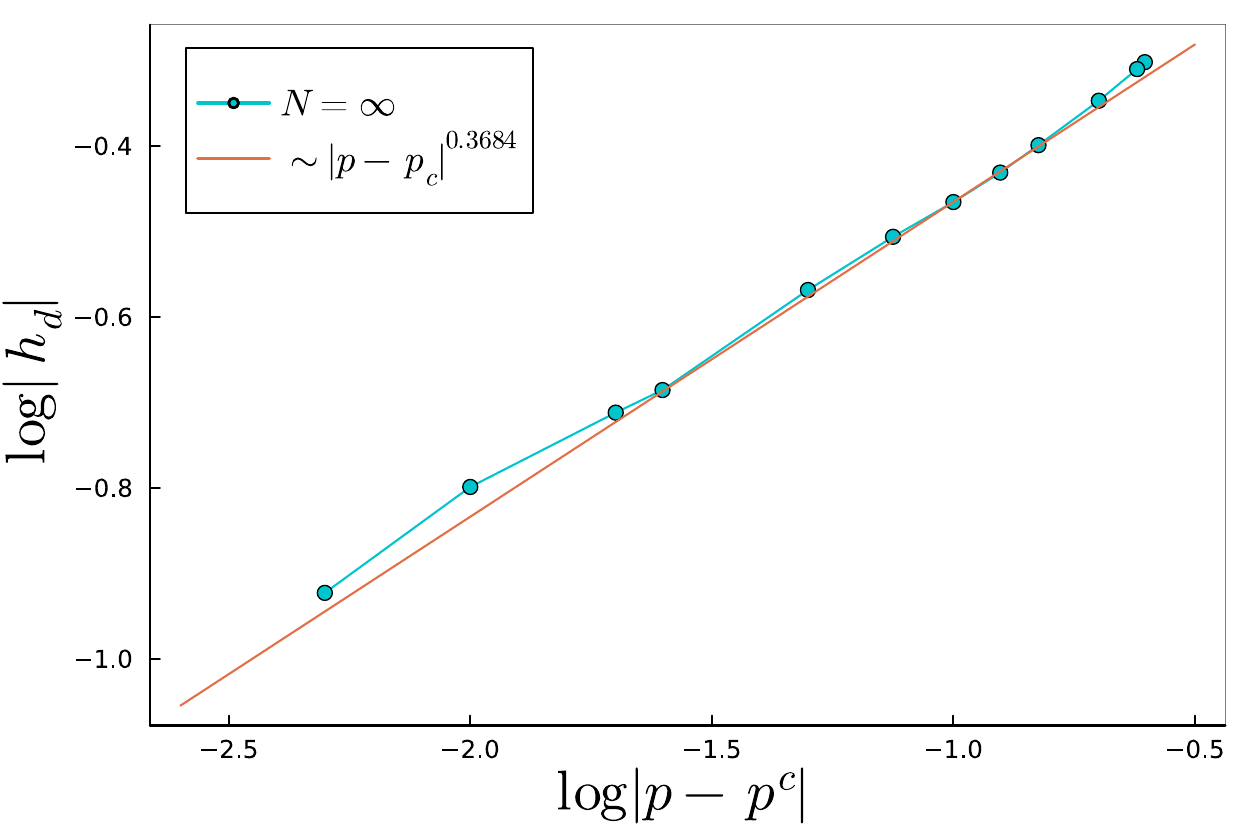}
    \caption{Fitting of the critical exponent $\beta$ from the extrapolated data of $h_d$ in the limit of $N \to \infty$.}
    \label{fig:beta_calc}
\end{figure}

Further analyzing the density of the Hamming distance $h_{d}$ as a function of the sampling parameter $p$ for increasing system size $N$, see Fig.~\ref{fig:Hamming_transition}, we observed a behavior that is reminiscent of the standard picture of the second order phase transition with critical parameter $p_c \approx 0.25$. While for $p<p_c$ $h_d$ seems to tend to a non-zero constant value as $N$ is increasing, for $p>p_{c}$ it is converging to zero. Near the critical point $p_c$, the Hamming distance density behaves as $h_{d} \sim (p_c-p)^{\beta}$ with the critical exponent $\beta \approx 0.368$ (see Fig.~\ref{fig:beta_calc}).

\section{The classical decorrelator and the correlation length}

We proceed with our phenomenological study of the phase transition by means of another dynamical quantity -- the decorrelator. This quantity allows us to define the analog of the correlation length and to study its divergence near the critical point, leading to the computation of another critical exponent.

\begin{figure}[ ]
    \centering
    \includegraphics[width=0.95\linewidth]{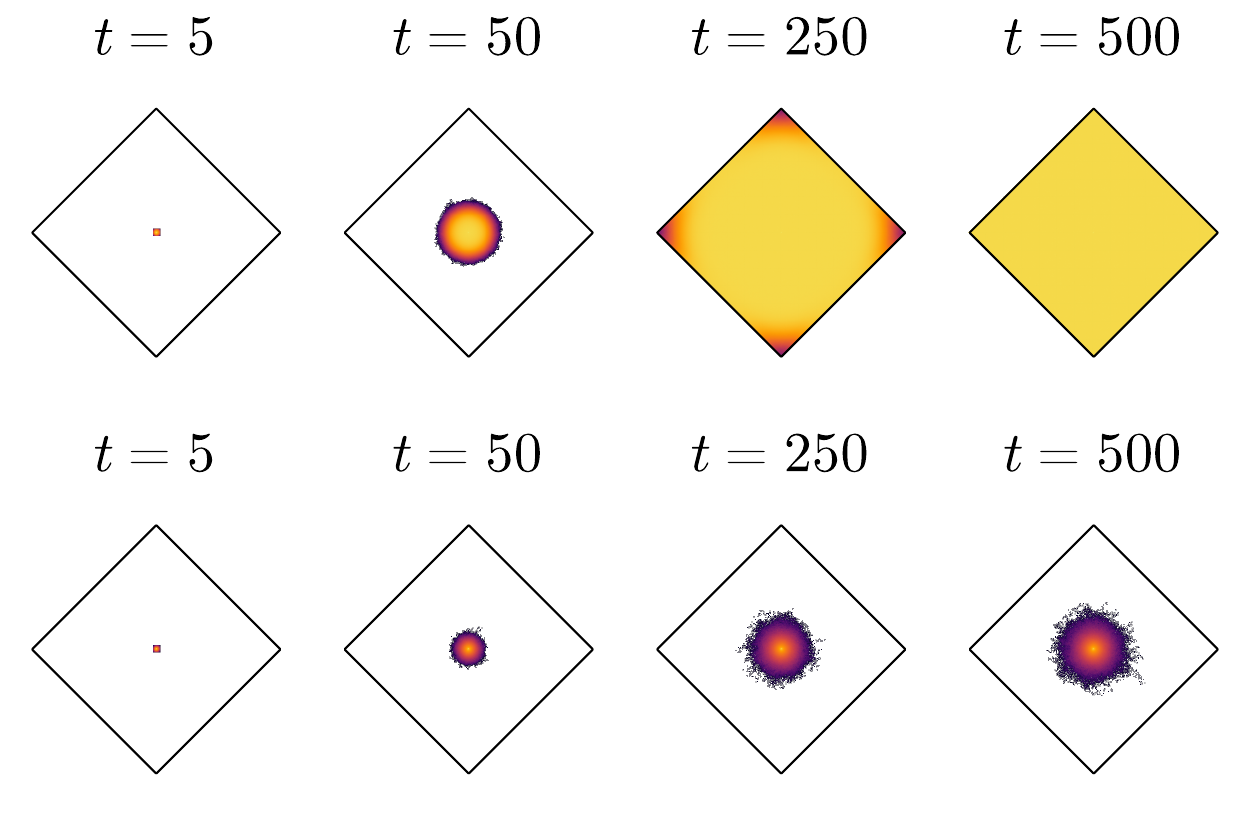}
    \caption{Example of the time evolution of the classical decorrelator for the square lattice of size $N=256$, the upper plots are for the system in the delocalized phase ($p=0.15$) while the lower plots are for the system in the localized phase ($p=0.35$). We note that we rotate the lattice by $45$ degrees for visualization purposes.}
    \label{fig:dec_time_evolutions}
\end{figure}

The decorrelator for initial configurations $\underline{s}^{A}$ and $\underline{s}^{B}$ is defined as:
\begin{equation}
    D (\mathbf{r} , t) \equiv \langle |s^{A}_{\mathbf{r}} (t) - s^{B}_{\mathbf{r}} (t) | \rangle\,,
\end{equation}
which can be interpreted as the specially resolved Hamming distance~\eqref{eq:hamming}. The bracket here again denotes an average over an ensemble of initial states with fixed $p$ (or fixed $\{n_q\}$). Due to spatial resolution we now 
consider $\underline{s}^{B}$ to differ from $\underline{s}^{A}$ by a bit flip at a fixed position $\mathbf{r}_0 = (0,0)$, implying the initial value 
\begin{equation}
    D (\mathbf{r} , 0) = \delta_{\mathbf{r} , \mathbf{0}}\,,
\end{equation}
indicating localization of the initial error.
The Hamming distance, in turn, is nothing but an integrated decorrelator: 
\begin{equation}
    H_{d} (t) = \sum_{\mathbf{r}} D(\mathbf{r} ,t)\,.
\end{equation} 
For later times, the error spreads out, and the support of the decorrelator typically grows, see Fig.~\ref{fig:dec_time_evolutions}. 

\begin{figure}[ ]
    \centering
    \includegraphics[width=0.95\linewidth]{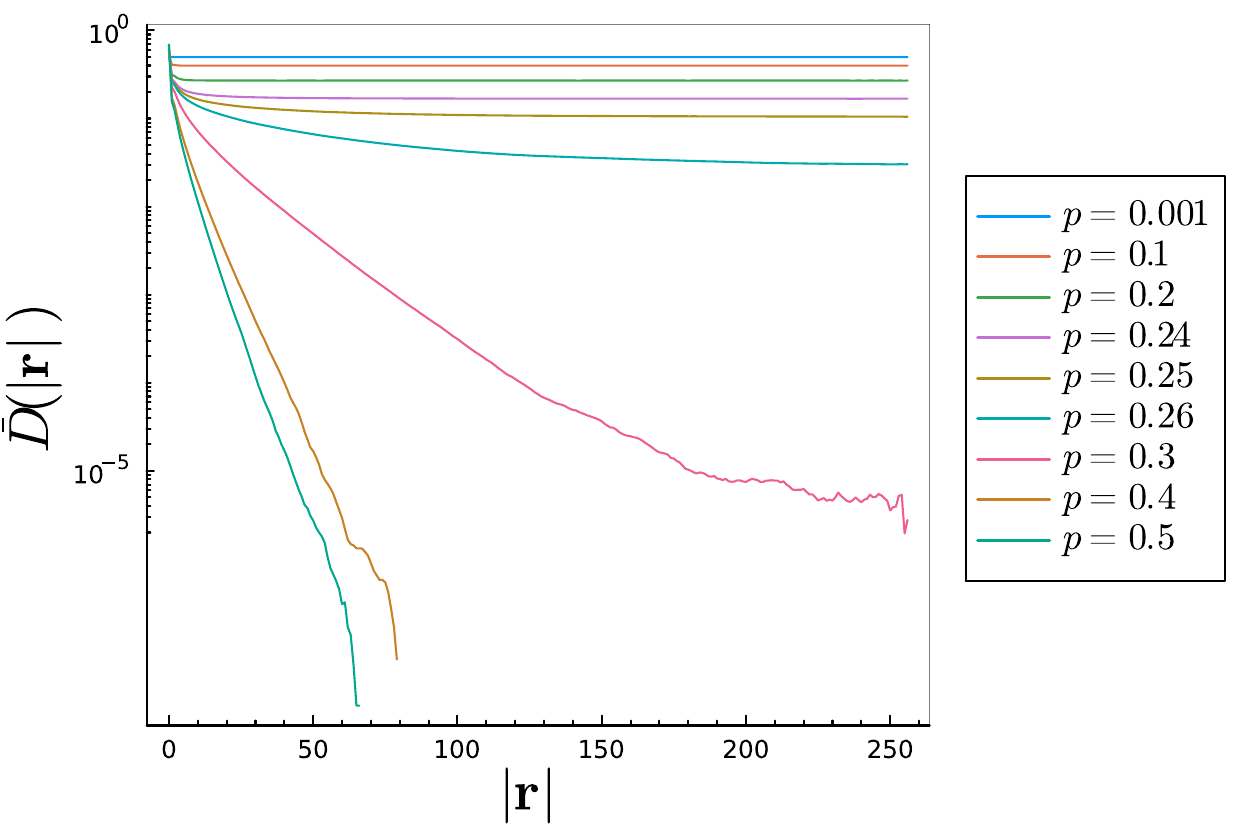}
    \caption{The time-averaged decorrelator $\bar{D}(\mathbf{r})$ as a function of the distance $|\mathbf{r}|$ for different values of the sampling probability $p$ of initial state. The data shown is from simulations of the square lattice of size $N=256\times256$ up to times $T=8000$ with $\mathcal{N}\approx 4\cdot 10^4$ Monte-Carlo samples.}
    \label{fig:dec_decay}
\end{figure}

Similarly, we will be interested in the long-time behavior of this quantity
\begin{equation}\label{time-aver-decorr}
    \bar{D}(\mathbf{r}) = \lim_{T\rightarrow\infty} \frac{1}{T} \sum_{T=0}^{t-1} D(\mathbf{r},t).
\end{equation}
In the delocalized phase, where the Hamming distance density $h_{d}$ is non-zero, one can expect that as a function of $\mathbf{r}$ the time-averaged decorrelator $\bar{D}(\mathbf{r})$ will behave approximately as a constant (this constant will then coincide with $h_{p}$ as $\bar{H}_d = \sum_{\mathbf{r}} \bar{D} (\mathbf{r} ) $). On the other hand, in the localized phase, the information about the error should spread only within a finite distance from the initial point. And so, $\bar{D}(\mathbf{r})$ should decay with the distance from the origin $|\mathbf{r}|$. Such behavior indeed takes place, as shown in Fig.~\ref{fig:dec_decay}.  

\begin{figure}[ ]
    \centering
    \includegraphics[width=0.9\linewidth]{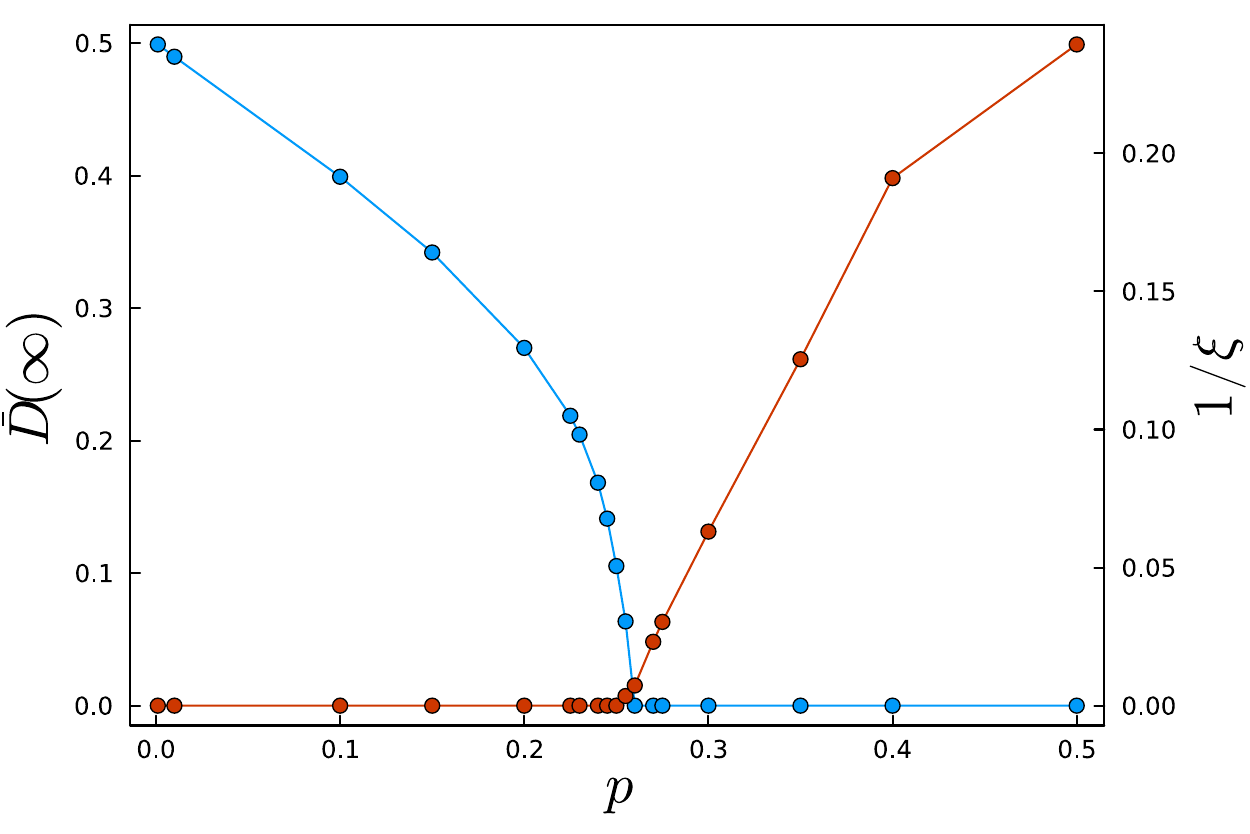}
    \caption{The plateau of the decorrelator $\bar{D}(\infty)$ as a function of the probability $p$ in blue (left axis) and the decay rate $1/\xi$ in red (right axis).
    }
    \label{fig:transition_point}
\end{figure}

Moreover, Fig.~\ref{fig:dec_decay} shows that in the localized phase the decorrelator decays exponentially, $D(\mathbf{r}) \sim e^{-|\mathbf{r}|/\xi}$, which allows us to define the correlation length $\xi$. In this regime $\xi$ remains finite, but it grows rapidly as the system approaches the critical point. At $p \to p_c+0$ the correlation length diverges, and it stays infinite throughout the delocalized phase (see Fig.~\ref{fig:transition_point}).

\begin{figure}[ ]
    \centering
    \includegraphics[width=0.9\linewidth]{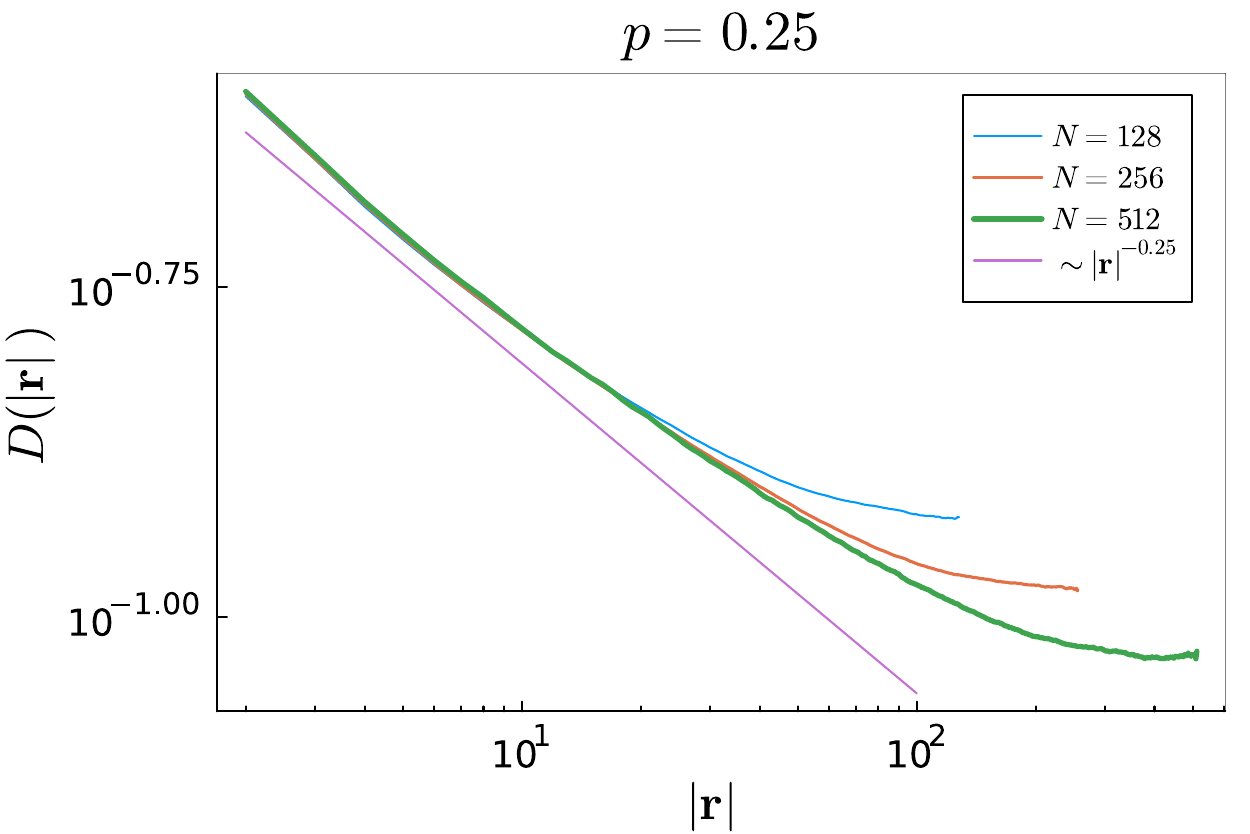}
    \caption{The power law decay of the decorrelator as a function of the distance $|\mathbf{r}|$ at the critical point $p_c=0.25$ for different system sizes $N\times N$. We plot the best power law fit in magenta.}
    \label{fig:power_law_deccay_decorr}
\end{figure}

Near the critical point, the correlation length diverges as $\xi \sim (p - p_c)^{-\nu}$, with an exponent $\nu \approx 1$. At the transition $p = p_c$, the decorrelator no longer decays exponentially but instead follows a power law, $\bar{D}(\mathbf{r}) \sim |\mathbf{r}|^{-\eta}$, with $\eta \approx 0.25$ (see Fig.~\ref{fig:power_law_deccay_decorr}).

\section{On the multifractality of dynamical structure factor}

In the previous sections, we showed the effects of the symmetries of the MCPCA and its initial state on its dynamics. Specifically, how changing the densities of the plaquette charges leads to a dynamical phase transition from localized to delocalized phase of information spreading. In this section, we show how these symmetries influence the multifractal behavior of the correlation function that was previously reported in~\cite{Kasim2025}. In this previous work, the effect of the symmetries has not been considered. The initial states were sampled uniformly from the maximum entropy ensemble $p=1/2$.

Firstly, we study how the behavior of the correlation function changes in different charge sectors, showing that the multifractal dynamical response, reported in~\cite{Kasim2025}, is becoming more and more smooth by decreasing the sampling parameter $p$. After that, we illustrate that the multifractality of the correlation function is related to dynamical localization of information observed here, and to the appearance of the local time-periodicities in the system.  

\subsection{The density-density correlation function across the transition}

On each vertex $v$ of the square lattice, we define a local observable, i.e. shifted particle density $n_v = \frac{1}{4} \sum_{e\in v} Z_{e}$. Then, the two-point correlation function averaged over an ensemble of states is defined as
\begin{equation}
    C (\mathbf{r}, t) = \langle n_{v+\mathbf{r}} (\Phi^{t}[\underline{s}]) n_v(\underline{s}) \rangle.
\end{equation}

\begin{figure}[ ]
    \centering
    \includegraphics[width=\linewidth]{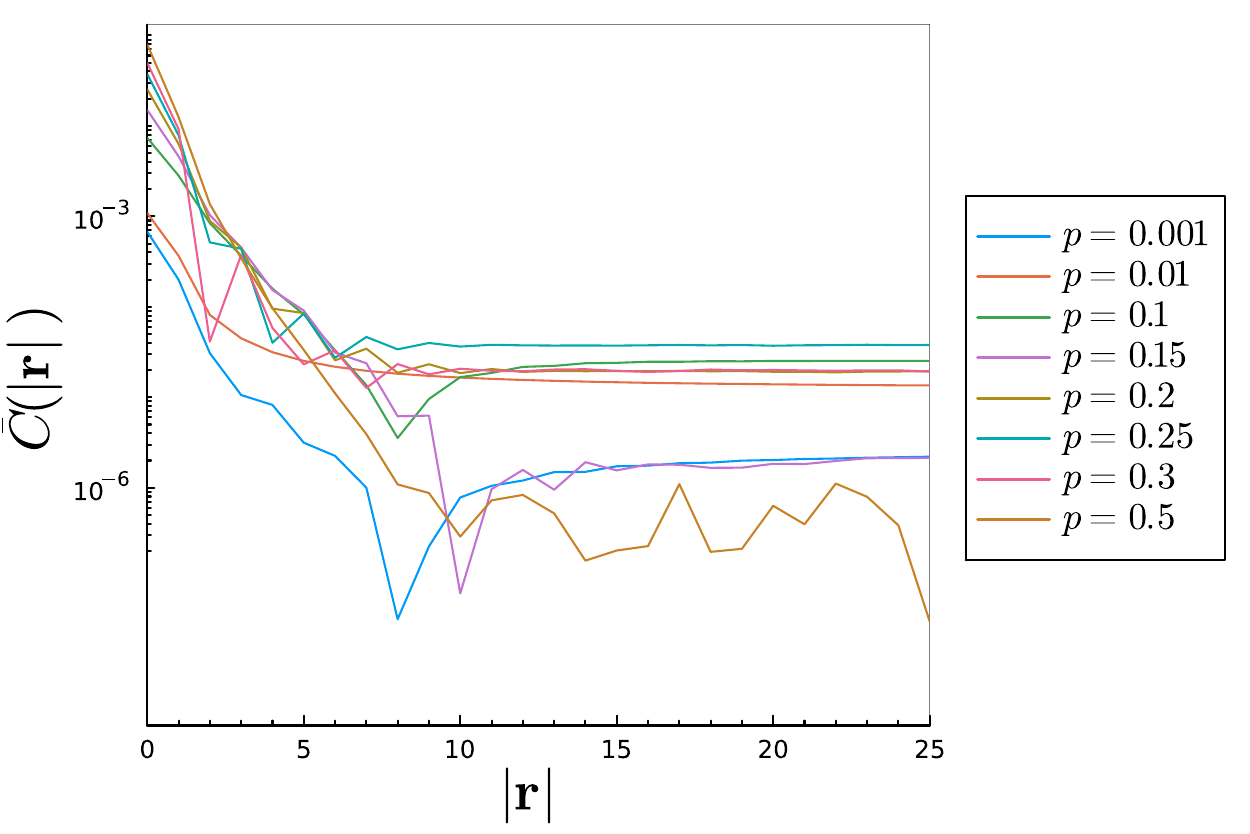}
    \caption{Time-averaged correlation function as a function of the distance $|\mathbf{r}|$ and for different sampling probabilities $p$.}
    \label{fig:correlator_r}
\end{figure}

We start our analysis by studying the time-averaged correlation function $\bar{C}(\mathbf{r}) = \lim_{T \rightarrow \infty} \frac{1}{T} \sum_{t=0}^{T-1} C(\mathbf{r} , t)$ in order to check if the properties of this quantity are similar to the decorrelator~(\ref{time-aver-decorr}). It turns out that for all values of $p$ the behavior of the correlator is qualitatively the same: initial exponential decay with the saturation at bigger values of $|\mathbf{r}|$, see Fig.\ref{fig:correlator_r}. Thus, the time-averaged 2-point correlation function is not sensitive to the phase transition in contrast with dynamics of the Hamming distance, Fig.~\ref{fig:Hamming_distance_different_NS}, and the decorrelator, Fig.~\ref{fig:dec_decay}. 

\begin{figure*}
    \centering
    \includegraphics[width=1\linewidth]{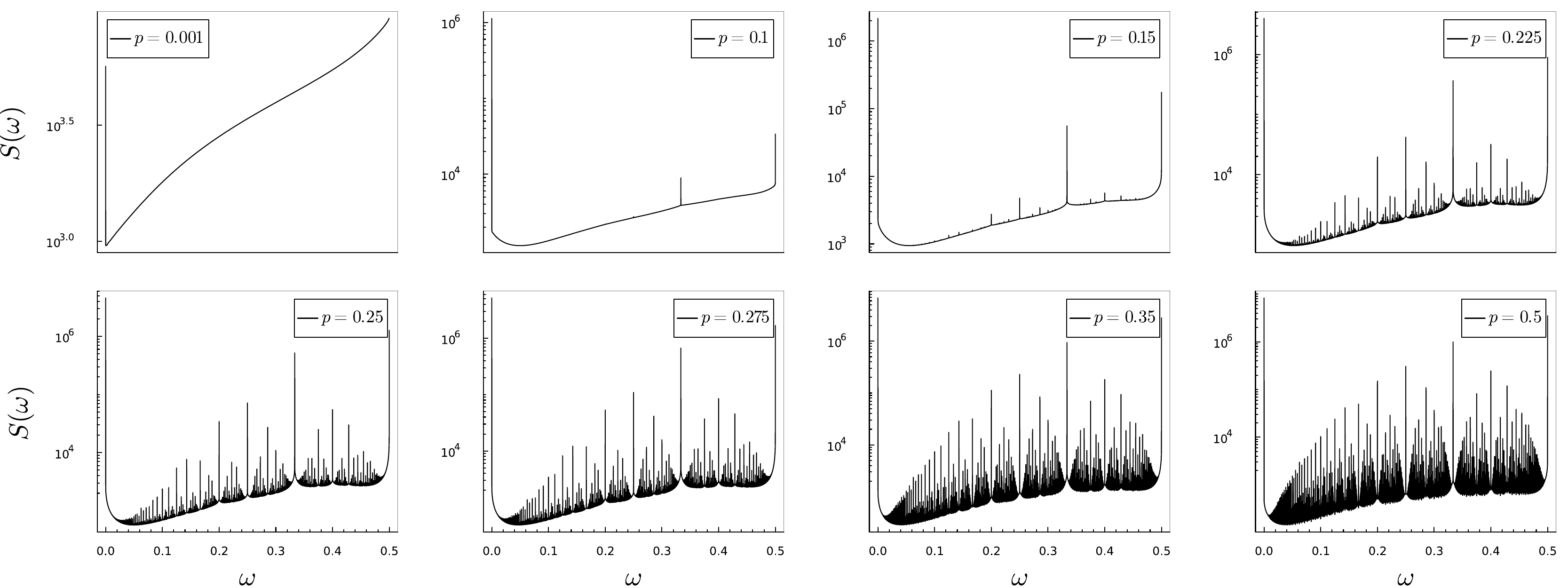}
    \caption{Power spectra $S(\omega)$ of the local density-density correlation function for system size $N=64\times 64$, the time signal was calculated for final time $T=2^{15}$ and for different sampling probability $p$ (different panels).}
    \label{fig:spectra_ni}
\end{figure*}

In addition, we explore the dynamical structure factor at $\mathbf{r} = 0$, i.e. the power spectrum
\begin{equation}
	S(\omega) =  \sum_{t\in \mathbb{Z}} C(\mathbf{r}=0,t) \exp(2\pi i t \omega).
\end{equation}
As we can see in Fig.~\ref{fig:spectra_ni}, the multifractal response of the correlation function that was observed in~\cite{Kasim2025} appears continuously in $p$, when the sampling parameter $p$ varies from $0$ to $0.5$, even though sub-harmonic response delta-spikes become weaker/sparser for smaller $p$. Thus, similarly to the time-averaged correlation function, even the multifractality of the 2-point function $S(\omega)$ cannot be an indicator of the phase transition.

\subsection{Local periodicities and the multifractality}

As we showed in previous sections, in the localized phase, the local error in the initial configuration can propagate only up to sub-extensive distances. For instance, for the case of $p=0.5$, the saturated value of the Hamming distance is almost independent of the system size. This hints at the fact that, even though the global period, i.e. Poincar\' e recurrence time, of the initial states of the system is usually exponential in the system size $N$, different local regions, in principle, can have much lower approximate periodicities. For example, a region in the system accidentally isolated with a closed loop that is in the Néel configuration (maximal loop charge) will evolve without any interaction with the rest of the system, resulting in a motion with a period that is much smaller than the global one. But even in less exceptional cases, the restrictions imposed by the  loop charges are decreasing the effective interaction between different regions of the system. As a result, different subsystems can behave approximately periodically with small periodicities (of order one).

\begin{figure}[ ]
    \centering
    \includegraphics[width=0.95\linewidth]{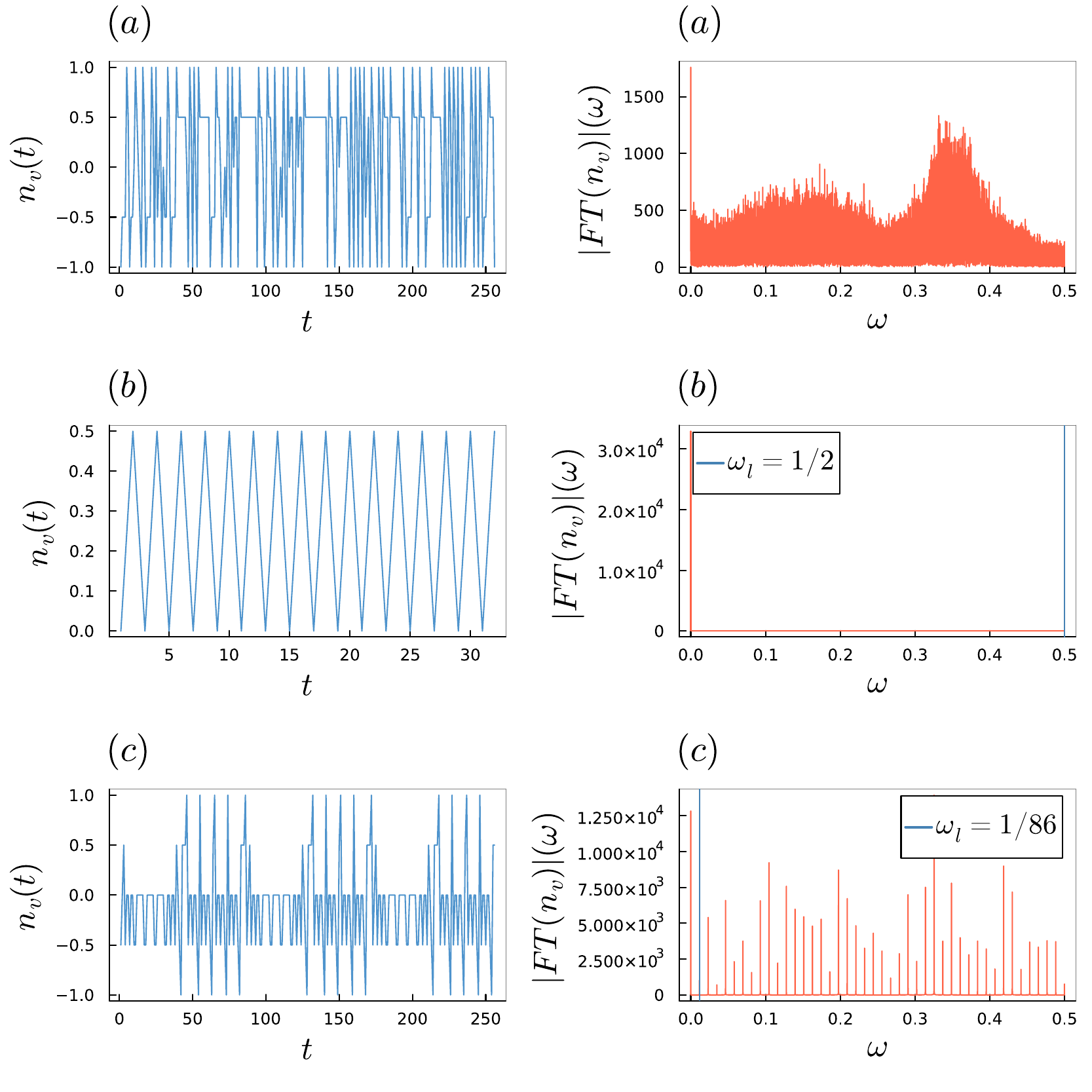}
    \caption{Example of the calculation of the local frequencies for three different local vertices $v$. On the left panels we show the real-time signal $n_v(t)$, and on the right panels we show the Fourier transform of this signal, all for the same -- typical initial configuration $\underline{s}$ for a lattice of size $N=32 \times 32$. (a) a local observable with a period that we could not resolve -- perhaps supporting the continuous part of the spectrum, (b) a signal with a simple local frequency $\omega_l=1/2$, (c) a more complex local frequency $\omega_l=1/86$ (notice that the peaks are equidistant from one another).}
    \label{fig:local_frequency_example}
\end{figure}

For a practical illustration, to calculate the local periods of the MCPCA for a specific sampling probability $p$, we first sample a random initial configuration $\underline{s}$, and then let the system evolve until time of $t=2N$ steps to ensure effective equilibration. After that, we evolve the system again up to times $T_f=2^{15}$ and record the evolution of the local density observable $n_v(t)$ on each vertex $v$ of the array.  
If the Fourier transform of the time evolution of the signal $n_v(t)$ has a prominent peak (or a series of equidistant peaks), we assign to it a local frequency (and its harmonic multiplicities), see Fig.\ref{fig:local_frequency_example}. By this method, we are able to reliably resolve periods up to $T_{\text{max}} \approx 150$.

\begin{figure}[ ]
    \centering
    \includegraphics[width=\linewidth]{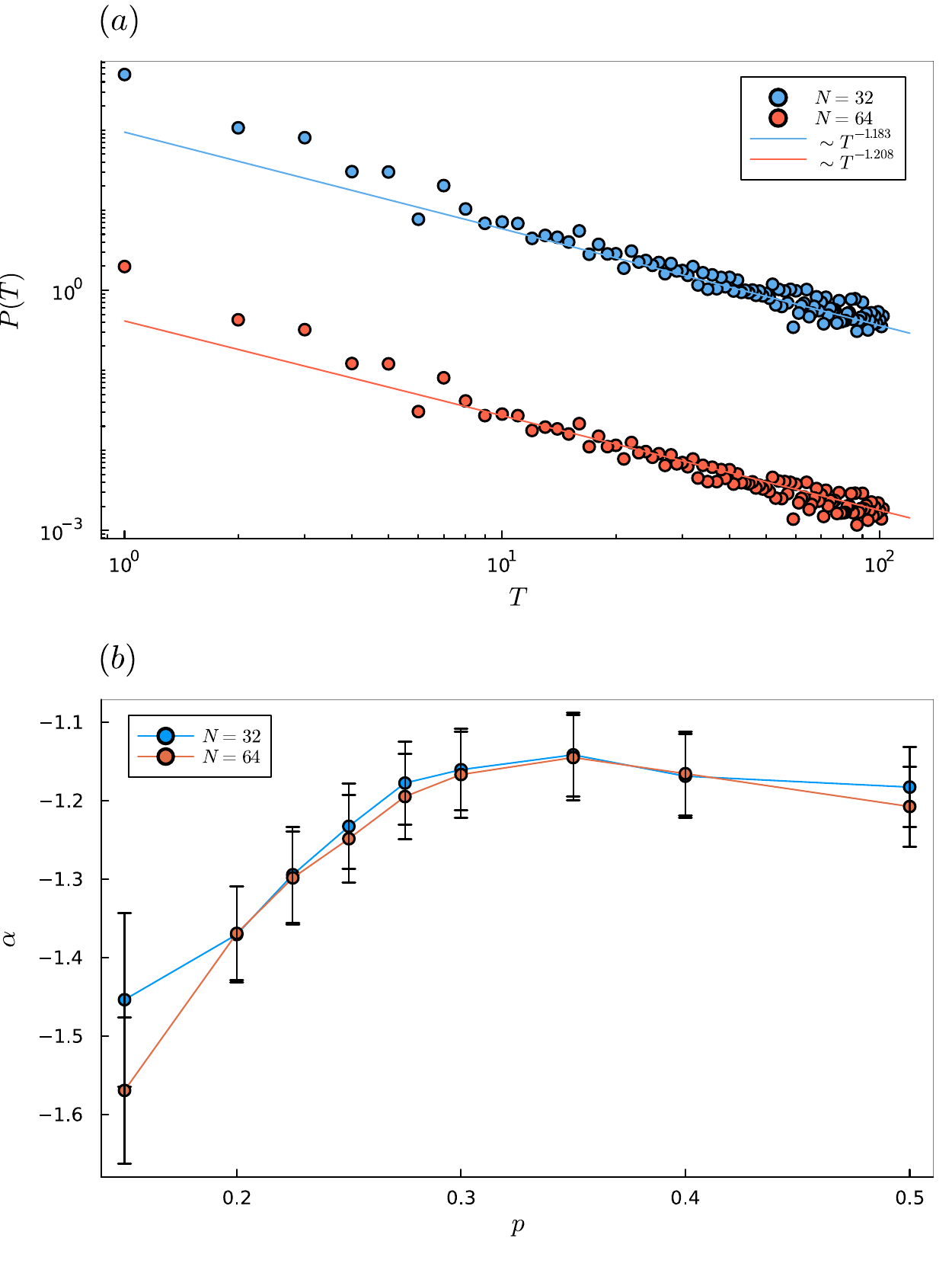}
    \caption{The calculation of the decay exponent $\alpha$ of the local period distribution. Panel (a) shows the distribution of local periods for a fixed sampling probability $p=0.5$ for different system sizes $N\times N$ and the best fit for the decay exponent $\alpha$. The plot for $N=64$ is shifted by a factor $10^{-3}$ for visualization purpose. Panel (b) shows the power law exponent of the distribution of local periods as a function of the sampling probability $p$. }
    \label{fig:alpha_calculation}
\end{figure}

\begin{figure}[ ]
    \centering
    \includegraphics[width=\linewidth]{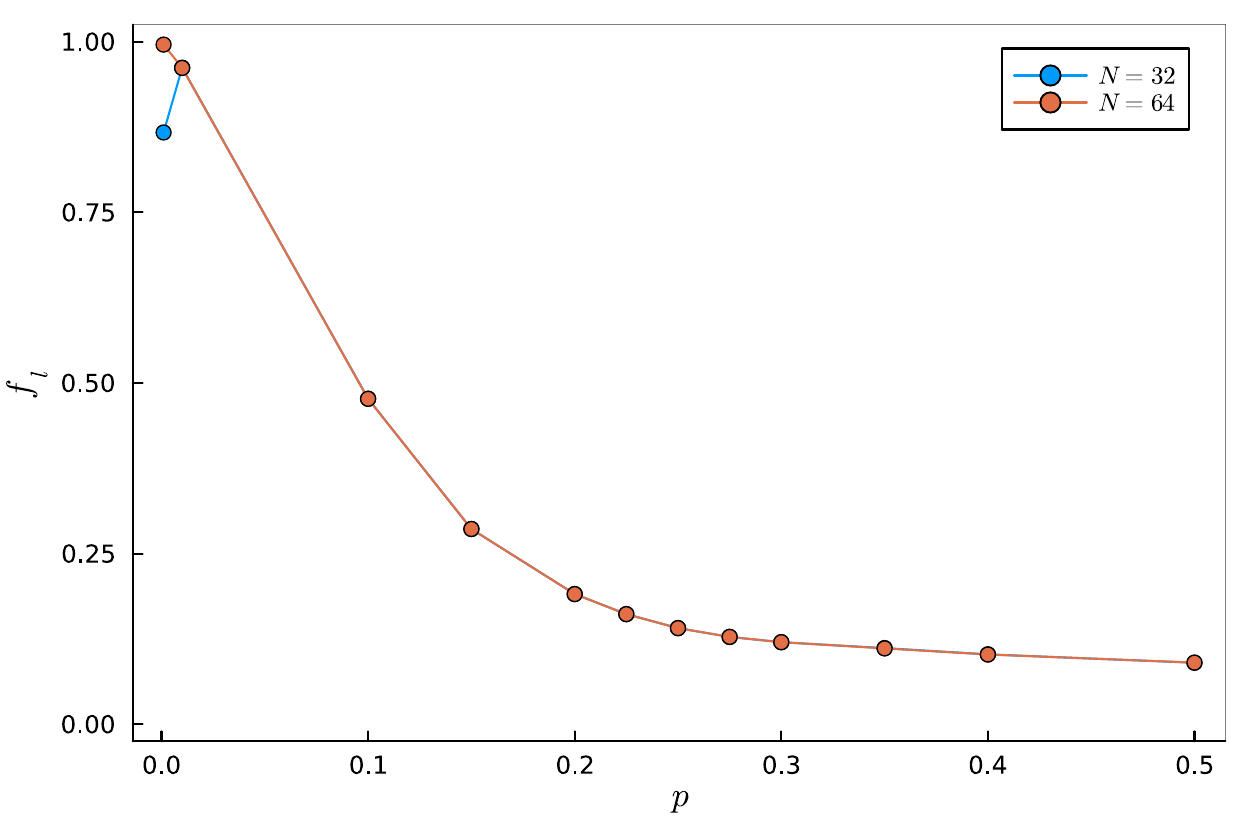}
    \caption{Fraction of the vertices with an unresolved period ($T>150$) $f_l$ as a function of the sampling probability $p$ for  two different system sizes $N\times N$.}
    \label{fig:fraction_unresolved}
\end{figure}

We have found that the distribution of local periods $P(T)$ decays with $T$ as a power law
\begin{equation}\label{distribution_periods}
    P(T) \sim T^{-\alpha}
\end{equation}
with the exponent $\alpha$ that depends on the sampling parameter $p$, see Fig.\ref{fig:alpha_calculation}. We also note here that the fraction of vertices with unresolved local period $T>T_{\text{max}}$ is smoothly decreasing as we increase the parameter $p$, since the dynamics is becoming more and more constrained, see Fig.~\ref{fig:fraction_unresolved}. This fraction should match with the total relative weight of the continuous part of the dynamical response function $S_{\rm reg}(\omega)$ as discussed in~\cite{Kasim2025}.
And again, this quantity is also not a good indicator of the phase transition discussed above.

However, the existence of local periods is directly related to the observation of the multifractality in MCPCA. Indeed, sampling a set of periods $\{ T_{i} \}_{i=1}^{\mathcal{N}}$ from the probability distribution~(\ref{distribution_periods}), one can assign to each period $T_i$ a uniform power-spectrum $S_{i} (\omega) = \frac{1}{T_i} \sum_{k=0}^{T_i-1} \delta ( \omega  - \frac{k}{T_i})$. Then, the average power-spectrum $\bar{S} (\omega) = \frac{1}{\mathcal{N}} \sum_{i} S_{i} (\omega) $ has qualitatively the same multifractal structure of peaks as the dynamical response function
in Fig.~\ref{fig:spectra_ni}.   

To find a quantitative similarity, one can consider the weights of peaks $A_{p,q}$ at rational frequencies $\omega = \frac{p}{q}$ of the power-spectrum $S(\omega)$ of the correlation function. In~\cite{Kasim2025}, it has been found that the total period-$q$ spectral weight $B_{q} = \sum_{p} A_{p,q}  $ has a power-law decay with $q$: $B_q \sim q^{-\mu}$. Assuming again our simplified picture of uniform spectral weights from all local periods, one can see that the total weight $B_{q}$ should be just proportional to the probability $P(T=q)$, and so the power $\alpha$ in Eq.(\ref{distribution_periods}) should coincide with the power $\mu$ of the decay of the spectral weight $B_q$. Both these exponents are indeed in very good agreement, showing that in our system the multifractality of the spectrum is a consequence of the presence of a multitude of local periodicities.

In contrast to the results of the probabilistic Kaufmann cellular automata~\cite{Weisbuch87}, we note no different behavior between the localized and delocalized information phases in respect to the local period. We see that the decay factor doesn't experience an important change as a function of $p$. We can then conclude that while the local periodicities are related to the observation of multifractality, they are not an indicator of the information spreading phase transition. Therefore, the two striking phenomena observed in MCPCA, namely the multifractal dynamical response, and dynamical phase transition in information spreading, can be physically unrelated, even though both arise as consequences of the symmetries of the model.

\section{Summary and conclusion}

We have discovered and analyzed a localization phase transition in the information spreading of deterministic momentum-conserving parity-check automata. These reversible automata belong to a class of systems that possess an extensive set of local conserved quantities --- 1-form symmetries --- which can be constructed as staggered-magnetization operators along all closed loops of the lattice \cite{PavelOrlov}. We demonstrated that, depending on the sector defined by these symmetries, the system dynamics can belong either to a delocalized or a localized phase of information spreading.

As an order parameter distinguishing these two phases, we employed the long-time limit of the density of the Hamming distance between two many-body trajectories that initially differ by a single cell. In the delocalized phase, a local error in the initial state propagates to an extensive scale, resulting in a finite Hamming-distance density. Conversely, in the localized phase, the error spreads only to a subextensive scale, leading to a vanishing density in the thermodynamic limit. Our numerical analysis indicates that these two regimes are separated by a second-order phase transition: the Hamming-distance density exhibits behavior analogous to the magnetization density in ferromagnetic–paramagnetic transitions. This transition is further characterized by the divergence of the correlation length, defined as the exponential decay length of the classical decorrelator. A summary of the scaling behavior across the different regimes is presented in Table~\ref{tab:scaling_summary}.

We also examined the multifractality of the power spectrum of the two-point correlation function, first observed in~\cite{Kasim2025}. Our findings show that the multifractal behavior of the spectrum is insensitive to the phase transition: the multifractal fraction evolves smoothly across the critical point. Nevertheless, we established a connection between multifractality and the emergence of approximate local periodicities in the system. These local periods arise from symmetry constraints, which are more pronounced in the localized phase. Moreover, we observed that the distribution of local periods follows a power-law form (\ref{distribution_periods}), consistent with the decay of the total period-$q$ spectral weight $B_q$ introduced in~\cite{Kasim2025}. It is worth noting that we observe this phase transition only in the square lattice. For MCPCA on the honeycomb lattice (as in~\cite{Kasim2025}), we find no phase transition, and the system remains delocalized in terms of information spreading. Further details on the honeycomb case are given in app.~\ref{apx:honeycomb}.

\renewcommand{\arraystretch}{1.3}
\begin{table}[H]
\centering
\begin{tabular}{|c|>{\centering\arraybackslash}p{1.8cm}|>{\centering\arraybackslash}p{1.4cm}|c|}
\hline
 & $h_d$ 
 & $\bar{D}(\mathbf{r})$ 
 & $\xi$ \\
\hline
 $p<p_c$ 
& $\sim (p_c - p)^{0.37}$ 
& const 
& $\infty$ \\
\hline
$p=p_c$ 
& $0$ 
& $\sim |\mathbf{r}|^{-0.25}$ 
& $\infty$ \\
\hline
$p>p_c$ 
& $0$ 
& $\sim e^{-|\mathbf{r}|/\xi}$ 
& $\sim (p-p_c)^{-1}$ \\
\hline
\end{tabular}
\caption{Scaling behavior of the Hamming-distance density $h_d$, the time-averaged decorrelator $\bar{D}(\mathbf{r})$, and the correlation length $\xi$ across the delocalized ($p<p_c$), critical ($p=p_c$), and localized ($p>p_c$) phases. The estimated value of the critical sampling parameter is $p_c \approx 0.25$. }
\label{tab:scaling_summary}
\end{table}

Our study opens several promising directions for future research. First, much of our analysis is numerical, and it remains important to clarify to what extent the observed phase transition can be described analytically, for example using renormalization-group techniques. In particular, computing the critical exponents and identifying the universality class of this transition is a straightforward open problem. Second, classical MCPCA represents a special case of a broader family of quantum circuits constructed from staggered-magnetization-conserving gates~\cite{PavelOrlov}. Extending the investigation to the quantum domain would therefore be a natural and intriguing next step. Finally, we have not explored the transport properties of the system. Since the two topological charges of MCPCA are extensive local conserved quantities, it would be of considerable interest to study their transport behavior across the different phases.

\section*{Acknowledgments}
We thank Cheryne Jonay, Katja Klobas and Pavel Kos for the fruitful discussions.
PO gratefully acknowledges Ferme de Champdolent and Didier Saint-Roch for providing an inspiring environment during the summer, where part of this manuscript was written.
This research has received funding from the European Union’s Horizon 2020 research and innovation program under the Marie Sklodowska-Curie grant agreement number 955479, as well as from
the European Research Council (ERC) through the Advanced grant QUEST (Grant Agreement No. 101096208) and from the Slovenian Research and Innovation agency (ARIS) through the program P1-0402 and grants N1-0219, N1-0368.

\appendix

\section{Honeycomb lattice}
\label{apx:honeycomb}
\begin{figure}[H]
    \centering
    \includegraphics[width=0.9\linewidth]{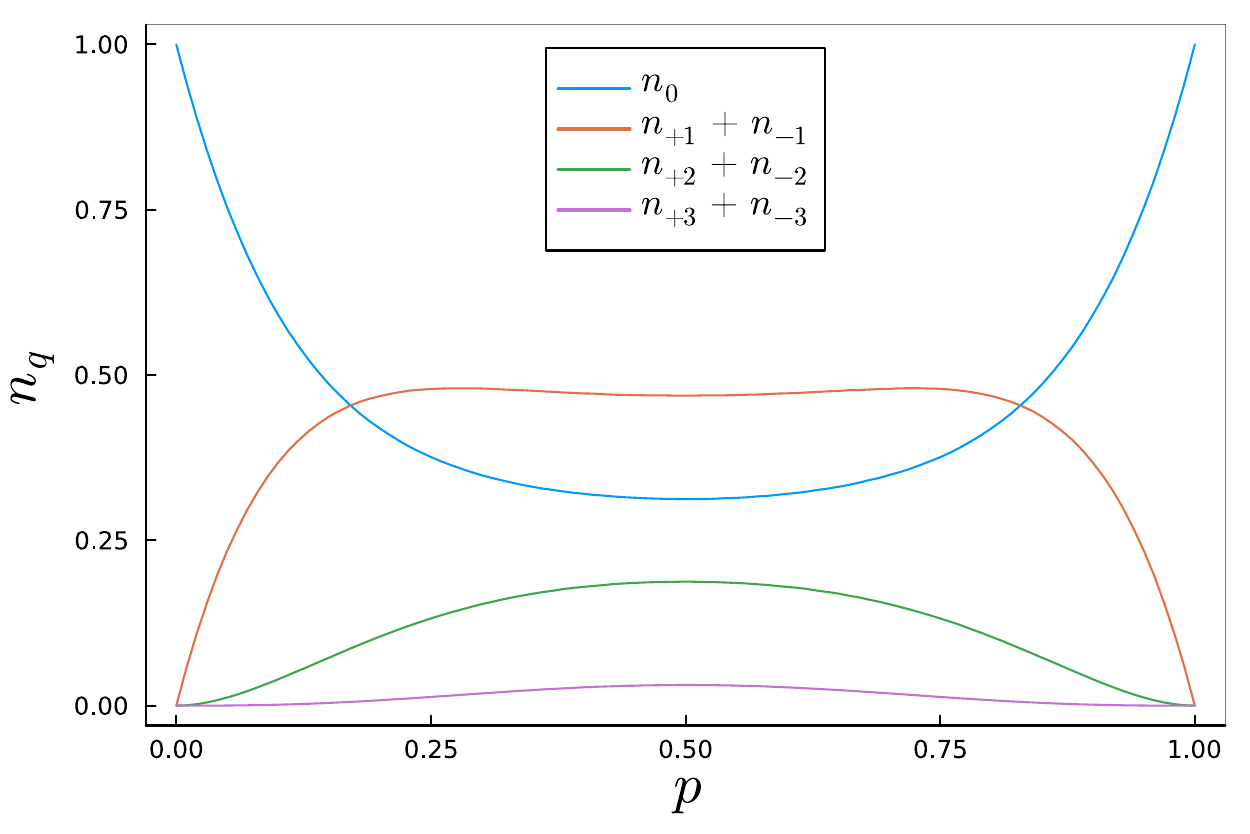}
    \caption{The density of plaquettes $n_q$ as a function of the probability of sampling $p$ for the honeycomb lattice.}
    \label{fig:sampling_probability_hex}
\end{figure}

We show the effect of changing the sampling probability on the honeycomb lattice. It is important to note that in the honeycomb lattice, the smallest plaquette is the following:
\begin{eqnarray}
\begin{tikzpicture}[baseline={(current bounding box.center)},every node/.style={inner sep=0,outer sep=0},line cap=rect,scale=0.7]
	\node (n0) at (0,0)[circle,fill,inner sep=1.25pt] {};
    \node (n1) at (0,0.5) {};
    \node (n2) at (-0.8660254037844387,-0.5)[circle,draw,inner sep=1.25pt] {};
	\node (n3) at (0.8660254037844387,-0.5)[circle,draw,inner sep=1.25pt] {};
    \node (n4) at (0.8660254037844387,-1.5)[circle,fill,inner sep=1.25pt] {};
	\node (n5) at (-0.8660254037844387,-1.5)[circle,fill,inner sep=1.25pt] {};
	\node (n6) at (0,-2)[circle,draw,inner sep=1.25pt] {};
	\node (n7) at (1.299038105676658,-0.25) {};
	\node (n8) at (-1.299038105676658,-0.25) {};
	\node (n9) at (1.299038105676658,-1.75) {};
	\node (n10) at (-1.299038105676658,-1.75) {};
	\node (n11) at (0,-2.5) {};
    \draw[-] (n0) -- (n1);
    \draw[-] (n0) -- (n2);
    \draw[-] (n0) -- (n3);
    \draw[-] (n3) -- (n4);
    \draw[-] (n2) -- (n5);
    \draw[-] (n5) -- (n6);
    \draw[-] (n4) -- (n6);
    \draw[-] (n6) -- (n11);
    \draw[-] (n4) -- (n9);
    \draw[-] (n5) -- (n10);
    \draw[-] (n3) -- (n7);
    \draw[-] (n2) -- (n8);
    \node[scale=1.8] (ar) at (0,-1) {$\circlearrowleft$};
	\draw[black,fill=red!50] (-0.43301270189221935cm+3.535533906pt,-0.25cm+3.535533906pt) arc  (45:225:5pt); 
	\draw[black,fill=white] (-0.43301270189221935cm-3.535533906pt,-0.25cm-3.535533906pt) arc  (225:405:5pt);
	\draw[black,fill=red!50] (0.43301270189221935cm+3.535533906pt,-0.25cm+3.535533906pt) arc  (45:225:5pt); 
	\draw[black,fill=white] (0.43301270189221935cm-3.535533906pt,-0.25cm-3.535533906pt) arc  (225:405:5pt);
	\draw[black,fill=red!50] (-0.8660254037844387cm+3.535533906pt,-1cm+3.535533906pt) arc  (45:225:5pt); 
	\draw[black,fill=white] (-0.8660254037844387cm-3.535533906pt,-1cm-3.535533906pt) arc  (225:405:5pt);
	\draw[black,fill=red!50] (0.8660254037844387cm+3.535533906pt,-1cm+3.535533906pt) arc  (45:225:5pt); 
	\draw[black,fill=white] (0.8660254037844387cm-3.535533906pt,-1cm-3.535533906pt) arc  (225:405:5pt);
	\draw[black,fill=red!50] (-0.43301270189221935cm+3.535533906pt,-1.75cm+3.535533906pt) arc  (45:225:5pt); 
	\draw[black,fill=white] (-0.43301270189221935cm-3.535533906pt,-1.75cm-3.535533906pt) arc  (225:405:5pt);
	\draw[black,fill=red!50] (0.43301270189221935cm+3.535533906pt,-1.75cm+3.535533906pt) arc  (45:225:5pt); 
	\draw[black,fill=white] (0.43301270189221935cm-3.535533906pt,-1.75cm-3.535533906pt) arc  (225:405:5pt);
	\end{tikzpicture}
	\label{eq:parity_honeycomb}
\end{eqnarray}
The plaquettes of the honeycomb lattice have $64$ possible configuration with $7$ values for the staggered magnetization instead of the $16$ configurations of the square lattice with $5$ possible values. We first show the relation of the density of plaquettes $n_q$ as a function of the sampling probability $p$ in fig~\ref{fig:sampling_probability_hex}. We note that the fraction of the frozen configurations ($M_\gamma = \pm3$) is way smaller than in  the square lattice. 

\begin{figure}[H]
    \centering
    \includegraphics[width=0.9\linewidth]{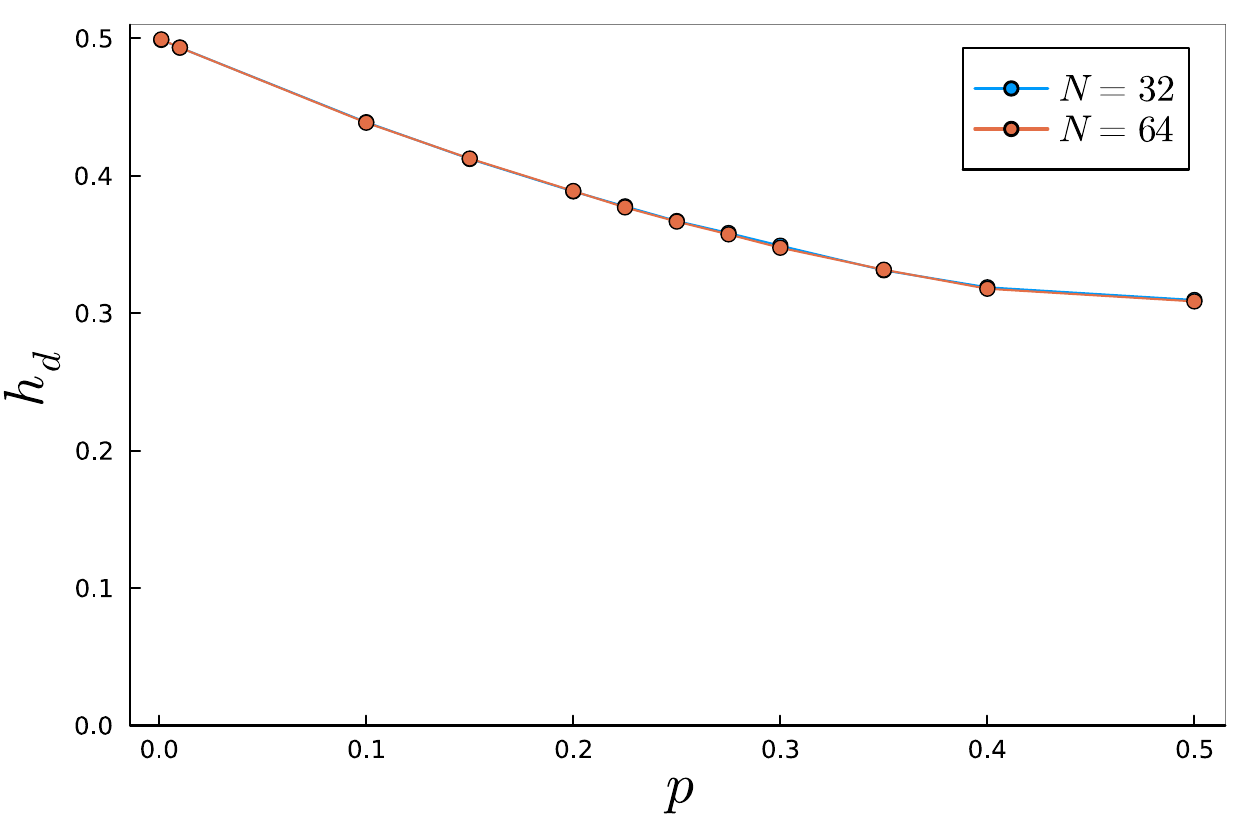}
    \caption{The density of the saturation values of the Hamming distance $h_d$ as a function of the sampling probability $p$ for the honeycomb lattice for different lattice sizes $N\times N$.}
    \label{fig:Hamming_hex}
\end{figure}

In Fig.~\ref{fig:Hamming_hex} we show the limit of $t\to \infty$ of the normalized Hamming distance. We see that the Hamming distance is lower for $p=0.5$ but it is still non-zero. The honeycomb lattice then doesn't exhibit a phase transition as the square lattice does. This observation may explain the fact that the multifractal regime in the Honeycomb lattice forms a smaller fraction of the total weight of the spectra than that of the square lattice~\cite{Kasim2025}.

\clearpage
\bibliography{bibli.bib}

\providecommand{\noopsort}[1]{}\providecommand{\singleletter}[1]{#1}%
\begin{thebibliography}{47}%
\makeatletter
\providecommand \@ifxundefined [1]{%
 \@ifx{#1\undefined}
}%
\providecommand \@ifnum [1]{%
 \ifnum #1\expandafter \@firstoftwo
 \else \expandafter \@secondoftwo
 \fi
}%
\providecommand \@ifx [1]{%
 \ifx #1\expandafter \@firstoftwo
 \else \expandafter \@secondoftwo
 \fi
}%
\providecommand \natexlab [1]{#1}%
\providecommand \enquote  [1]{``#1''}%
\providecommand \bibnamefont  [1]{#1}%
\providecommand \bibfnamefont [1]{#1}%
\providecommand \citenamefont [1]{#1}%
\providecommand \href@noop [0]{\@secondoftwo}%
\providecommand \href [0]{\begingroup \@sanitize@url \@href}%
\providecommand \@href[1]{\@@startlink{#1}\@@href}%
\providecommand \@@href[1]{\endgroup#1\@@endlink}%
\providecommand \@sanitize@url [0]{\catcode `\\12\catcode `\$12\catcode
  `\&12\catcode `\#12\catcode `\^12\catcode `\_12\catcode `\%12\relax}%
\providecommand \@@startlink[1]{}%
\providecommand \@@endlink[0]{}%
\providecommand \url  [0]{\begingroup\@sanitize@url \@url }%
\providecommand \@url [1]{\endgroup\@href {#1}{\urlprefix }}%
\providecommand \urlprefix  [0]{URL }%
\providecommand \Eprint [0]{\href }%
\providecommand \doibase [0]{https://doi.org/}%
\providecommand \selectlanguage [0]{\@gobble}%
\providecommand \bibinfo  [0]{\@secondoftwo}%
\providecommand \bibfield  [0]{\@secondoftwo}%
\providecommand \translation [1]{[#1]}%
\providecommand \BibitemOpen [0]{}%
\providecommand \bibitemStop [0]{}%
\providecommand \bibitemNoStop [0]{.\EOS\space}%
\providecommand \EOS [0]{\spacefactor3000\relax}%
\providecommand \BibitemShut  [1]{\csname bibitem#1\endcsname}%
\let\auto@bib@innerbib\@empty
\bibitem [{\citenamefont {Kasim}\ and\ \citenamefont
  {Prosen}(2025)}]{Kasim2025}%
  \BibitemOpen
  \bibfield  {author} {\bibinfo {author} {\bibfnamefont {Y.}~\bibnamefont
  {Kasim}}\ and\ \bibinfo {author} {\bibfnamefont {T.}~\bibnamefont {Prosen}},\
  }\bibfield  {title} {\bibinfo {title} {Deterministic many-body dynamics with
  multifractal response},\ }\href
  {https://doi.org/10.1103/PhysRevResearch.7.023230} {\bibfield  {journal}
  {\bibinfo  {journal} {Phys. Rev. Res.}\ }\textbf {\bibinfo {volume} {7}},\
  \bibinfo {pages} {023230} (\bibinfo {year} {2025})}\BibitemShut {NoStop}%
\bibitem [{\citenamefont {Arnold}\ \emph {et~al.}(2006)\citenamefont {Arnold},
  \citenamefont {Kozlov}, \citenamefont {Neishtadt},\ and\ \citenamefont
  {Iacob}}]{ArnoldBook}%
  \BibitemOpen
  \bibfield  {author} {\bibinfo {author} {\bibfnamefont {V.~I.}\ \bibnamefont
  {Arnold}}, \bibinfo {author} {\bibfnamefont {V.~V.}\ \bibnamefont {Kozlov}},
  \bibinfo {author} {\bibfnamefont {A.~I.}\ \bibnamefont {Neishtadt}},\ and\
  \bibinfo {author} {\bibfnamefont {I.}~\bibnamefont {Iacob}},\ }\href@noop {}
  {\emph {\bibinfo {title} {Mathematical aspects of classical and celestial
  mechanics}}},\ Vol.~\bibinfo {volume} {3}\ (\bibinfo  {publisher}
  {Springer},\ \bibinfo {year} {2006})\BibitemShut {NoStop}%
\bibitem [{\citenamefont {Fermi}\ \emph {et~al.}(1955)\citenamefont {Fermi},
  \citenamefont {Pasta}, \citenamefont {Ulam},\ and\ \citenamefont
  {Tsingou}}]{Fermi55}%
  \BibitemOpen
  \bibfield  {author} {\bibinfo {author} {\bibfnamefont {E.}~\bibnamefont
  {Fermi}}, \bibinfo {author} {\bibfnamefont {P.}~\bibnamefont {Pasta}},
  \bibinfo {author} {\bibfnamefont {S.}~\bibnamefont {Ulam}},\ and\ \bibinfo
  {author} {\bibfnamefont {M.}~\bibnamefont {Tsingou}},\ }\bibfield  {title}
  {\bibinfo {title} {Studies of the nonlinear problems},\ }\bibfield  {journal}
  {\bibinfo  {journal} {Technical report}\ }\href
  {https://doi.org/10.2172/4376203} {10.2172/4376203} (\bibinfo {year}
  {1955})\BibitemShut {NoStop}%
\bibitem [{\citenamefont {Berman}\ and\ \citenamefont
  {Izrailev}(2005)}]{Izrailev}%
  \BibitemOpen
  \bibfield  {author} {\bibinfo {author} {\bibfnamefont {G.}~\bibnamefont
  {Berman}}\ and\ \bibinfo {author} {\bibfnamefont {F.}~\bibnamefont
  {Izrailev}},\ }\bibfield  {title} {\bibinfo {title} {The fermi--pasta--ulam
  problem: fifty years of progress},\ }\href
  {https://doi.org/10.1063/1.1855036} {\bibfield  {journal} {\bibinfo
  {journal} {Chaos: An Interdisciplinary Journal of Nonlinear Science}\
  }\textbf {\bibinfo {volume} {15}} (\bibinfo {year} {2005})}\BibitemShut
  {NoStop}%
\bibitem [{\citenamefont {Prosen}(1998)}]{Prosen98}%
  \BibitemOpen
  \bibfield  {author} {\bibinfo {author} {\bibfnamefont {T.}~\bibnamefont
  {Prosen}},\ }\bibfield  {title} {\bibinfo {title} {Time evolution of a
  quantum many-body system: Transition from integrability to ergodicity in the
  thermodynamic limit},\ }\href {https://doi.org/10.1103/PhysRevLett.80.1808}
  {\bibfield  {journal} {\bibinfo  {journal} {Phys. Rev. Lett.}\ }\textbf
  {\bibinfo {volume} {80}},\ \bibinfo {pages} {1808} (\bibinfo {year}
  {1998})}\BibitemShut {NoStop}%
\bibitem [{\citenamefont {Prosen}(2007)}]{Prosen07}%
  \BibitemOpen
  \bibfield  {author} {\bibinfo {author} {\bibfnamefont {T.}~\bibnamefont
  {Prosen}},\ }\bibfield  {title} {\bibinfo {title} {Chaos and complexity of
  quantum motion},\ }\href {https://doi.org/10.1088/1751-8113/40/28/S02}
  {\bibfield  {journal} {\bibinfo  {journal} {Journal of Physics A:
  Mathematical and Theoretical}\ }\textbf {\bibinfo {volume} {40}},\ \bibinfo
  {pages} {7881} (\bibinfo {year} {2007})}\BibitemShut {NoStop}%
\bibitem [{\citenamefont {Maldacena}\ \emph {et~al.}(2016)\citenamefont
  {Maldacena}, \citenamefont {Shenker},\ and\ \citenamefont
  {Stanford}}]{Maldacena2016}%
  \BibitemOpen
  \bibfield  {author} {\bibinfo {author} {\bibfnamefont {J.}~\bibnamefont
  {Maldacena}}, \bibinfo {author} {\bibfnamefont {S.~H.}\ \bibnamefont
  {Shenker}},\ and\ \bibinfo {author} {\bibfnamefont {D.}~\bibnamefont
  {Stanford}},\ }\bibfield  {title} {\bibinfo {title} {A bound on chaos},\
  }\bibfield  {journal} {\bibinfo  {journal} {Journal of High Energy Physics}\
  }\textbf {\bibinfo {volume} {2016}},\ \href
  {https://doi.org/10.1007/jhep08(2016)106} {10.1007/jhep08(2016)106} (\bibinfo
  {year} {2016})\BibitemShut {NoStop}%
\bibitem [{\citenamefont {Hashimoto}\ \emph {et~al.}(2017)\citenamefont
  {Hashimoto}, \citenamefont {Murata},\ and\ \citenamefont
  {Yoshii}}]{Hashimoto2017}%
  \BibitemOpen
  \bibfield  {author} {\bibinfo {author} {\bibfnamefont {K.}~\bibnamefont
  {Hashimoto}}, \bibinfo {author} {\bibfnamefont {K.}~\bibnamefont {Murata}},\
  and\ \bibinfo {author} {\bibfnamefont {R.}~\bibnamefont {Yoshii}},\
  }\bibfield  {title} {\bibinfo {title} {Out-of-time-order correlators in
  quantum mechanics},\ }\bibfield  {journal} {\bibinfo  {journal} {Journal of
  High Energy Physics}\ }\textbf {\bibinfo {volume} {2017}},\ \href
  {https://doi.org/10.1007/jhep10(2017)138} {10.1007/jhep10(2017)138} (\bibinfo
  {year} {2017})\BibitemShut {NoStop}%
\bibitem [{\citenamefont {Swingle}(2018)}]{Swingle:2018ekw}%
  \BibitemOpen
  \bibfield  {author} {\bibinfo {author} {\bibfnamefont {B.}~\bibnamefont
  {Swingle}},\ }\bibfield  {title} {\bibinfo {title} {{Unscrambling the physics
  of out-of-time-order correlators}},\ }\href
  {https://doi.org/10.1038/s41567-018-0295-5} {\bibfield  {journal} {\bibinfo
  {journal} {Nature Phys.}\ }\textbf {\bibinfo {volume} {14}},\ \bibinfo
  {pages} {988} (\bibinfo {year} {2018})}\BibitemShut {NoStop}%
\bibitem [{\citenamefont {Bilitewski}\ \emph {et~al.}(2018)\citenamefont
  {Bilitewski}, \citenamefont {Bhattacharjee},\ and\ \citenamefont
  {Moessner}}]{Bilitewski2018}%
  \BibitemOpen
  \bibfield  {author} {\bibinfo {author} {\bibfnamefont {T.}~\bibnamefont
  {Bilitewski}}, \bibinfo {author} {\bibfnamefont {S.}~\bibnamefont
  {Bhattacharjee}},\ and\ \bibinfo {author} {\bibfnamefont {R.}~\bibnamefont
  {Moessner}},\ }\bibfield  {title} {\bibinfo {title} {Temperature dependence
  of the butterfly effect in a classical many-body system},\ }\href
  {https://doi.org/10.1103/PhysRevLett.121.250602} {\bibfield  {journal}
  {\bibinfo  {journal} {Phys. Rev. Lett.}\ }\textbf {\bibinfo {volume} {121}},\
  \bibinfo {pages} {250602} (\bibinfo {year} {2018})}\BibitemShut {NoStop}%
\bibitem [{\citenamefont {Das}\ \emph {et~al.}(2018)\citenamefont {Das},
  \citenamefont {Chakrabarty}, \citenamefont {Dhar}, \citenamefont {Kundu},
  \citenamefont {Huse}, \citenamefont {Moessner}, \citenamefont {Ray},\ and\
  \citenamefont {Bhattacharjee}}]{Das2018}%
  \BibitemOpen
  \bibfield  {author} {\bibinfo {author} {\bibfnamefont {A.}~\bibnamefont
  {Das}}, \bibinfo {author} {\bibfnamefont {S.}~\bibnamefont {Chakrabarty}},
  \bibinfo {author} {\bibfnamefont {A.}~\bibnamefont {Dhar}}, \bibinfo {author}
  {\bibfnamefont {A.}~\bibnamefont {Kundu}}, \bibinfo {author} {\bibfnamefont
  {D.~A.}\ \bibnamefont {Huse}}, \bibinfo {author} {\bibfnamefont
  {R.}~\bibnamefont {Moessner}}, \bibinfo {author} {\bibfnamefont {S.~S.}\
  \bibnamefont {Ray}},\ and\ \bibinfo {author} {\bibfnamefont {S.}~\bibnamefont
  {Bhattacharjee}},\ }\bibfield  {title} {\bibinfo {title} {Light-cone
  spreading of perturbations and the butterfly effect in a classical spin
  chain},\ }\href {https://doi.org/10.1103/PhysRevLett.121.024101} {\bibfield
  {journal} {\bibinfo  {journal} {Phys. Rev. Lett.}\ }\textbf {\bibinfo
  {volume} {121}},\ \bibinfo {pages} {024101} (\bibinfo {year}
  {2018})}\BibitemShut {NoStop}%
\bibitem [{\citenamefont {Bilitewski}\ \emph {et~al.}(2021)\citenamefont
  {Bilitewski}, \citenamefont {Bhattacharjee},\ and\ \citenamefont
  {Moessner}}]{Bilitewski2021}%
  \BibitemOpen
  \bibfield  {author} {\bibinfo {author} {\bibfnamefont {T.}~\bibnamefont
  {Bilitewski}}, \bibinfo {author} {\bibfnamefont {S.}~\bibnamefont
  {Bhattacharjee}},\ and\ \bibinfo {author} {\bibfnamefont {R.}~\bibnamefont
  {Moessner}},\ }\bibfield  {title} {\bibinfo {title} {Classical many-body
  chaos with and without quasiparticles},\ }\href
  {https://doi.org/10.1103/PhysRevB.103.174302} {\bibfield  {journal} {\bibinfo
   {journal} {Phys. Rev. B}\ }\textbf {\bibinfo {volume} {103}},\ \bibinfo
  {pages} {174302} (\bibinfo {year} {2021})}\BibitemShut {NoStop}%
\bibitem [{\citenamefont {Murugan}\ \emph {et~al.}(2021)\citenamefont
  {Murugan}, \citenamefont {Kumar}, \citenamefont {Bhattacharjee},\ and\
  \citenamefont {Ray}}]{Murugan2021}%
  \BibitemOpen
  \bibfield  {author} {\bibinfo {author} {\bibfnamefont {S.~D.}\ \bibnamefont
  {Murugan}}, \bibinfo {author} {\bibfnamefont {D.}~\bibnamefont {Kumar}},
  \bibinfo {author} {\bibfnamefont {S.}~\bibnamefont {Bhattacharjee}},\ and\
  \bibinfo {author} {\bibfnamefont {S.~S.}\ \bibnamefont {Ray}},\ }\bibfield
  {title} {\bibinfo {title} {Many-body chaos in thermalized fluids},\ }\href
  {https://doi.org/10.1103/PhysRevLett.127.124501} {\bibfield  {journal}
  {\bibinfo  {journal} {Phys. Rev. Lett.}\ }\textbf {\bibinfo {volume} {127}},\
  \bibinfo {pages} {124501} (\bibinfo {year} {2021})}\BibitemShut {NoStop}%
\bibitem [{\citenamefont {Liu}\ \emph {et~al.}(2021)\citenamefont {Liu},
  \citenamefont {Willsher}, \citenamefont {Bilitewski}, \citenamefont {Li},
  \citenamefont {Smith}, \citenamefont {Christensen}, \citenamefont
  {Moessner},\ and\ \citenamefont {Knolle}}]{Liu2021}%
  \BibitemOpen
  \bibfield  {author} {\bibinfo {author} {\bibfnamefont {S.-W.}\ \bibnamefont
  {Liu}}, \bibinfo {author} {\bibfnamefont {J.}~\bibnamefont {Willsher}},
  \bibinfo {author} {\bibfnamefont {T.}~\bibnamefont {Bilitewski}}, \bibinfo
  {author} {\bibfnamefont {J.-J.}\ \bibnamefont {Li}}, \bibinfo {author}
  {\bibfnamefont {A.}~\bibnamefont {Smith}}, \bibinfo {author} {\bibfnamefont
  {K.}~\bibnamefont {Christensen}}, \bibinfo {author} {\bibfnamefont
  {R.}~\bibnamefont {Moessner}},\ and\ \bibinfo {author} {\bibfnamefont
  {J.}~\bibnamefont {Knolle}},\ }\bibfield  {title} {\bibinfo {title}
  {Butterfly effect and spatial structure of information spreading in a chaotic
  cellular automaton},\ }\href {https://doi.org/10.1103/PhysRevB.103.094109}
  {\bibfield  {journal} {\bibinfo  {journal} {Phys. Rev. B}\ }\textbf {\bibinfo
  {volume} {103}},\ \bibinfo {pages} {094109} (\bibinfo {year}
  {2021})}\BibitemShut {NoStop}%
\bibitem [{\citenamefont {Bertini}\ \emph {et~al.}(2025)\citenamefont
  {Bertini}, \citenamefont {Klobas}, \citenamefont {Kos},\ and\ \citenamefont
  {Malz}}]{Bertini2025}%
  \BibitemOpen
  \bibfield  {author} {\bibinfo {author} {\bibfnamefont {B.}~\bibnamefont
  {Bertini}}, \bibinfo {author} {\bibfnamefont {K.}~\bibnamefont {Klobas}},
  \bibinfo {author} {\bibfnamefont {P.}~\bibnamefont {Kos}},\ and\ \bibinfo
  {author} {\bibfnamefont {D.}~\bibnamefont {Malz}},\ }\bibfield  {title}
  {\bibinfo {title} {Quantum and classical dynamics with random permutation
  circuits},\ }\href {https://doi.org/10.1103/PhysRevX.15.011015} {\bibfield
  {journal} {\bibinfo  {journal} {Phys. Rev. X}\ }\textbf {\bibinfo {volume}
  {15}},\ \bibinfo {pages} {011015} (\bibinfo {year} {2025})}\BibitemShut
  {NoStop}%
\bibitem [{\citenamefont {Hamming}(1950)}]{Hamming1950}%
  \BibitemOpen
  \bibfield  {author} {\bibinfo {author} {\bibfnamefont {R.~W.}\ \bibnamefont
  {Hamming}},\ }\bibfield  {title} {\bibinfo {title} {Error detecting and error
  correcting codes},\ }\href
  {https://doi.org/10.1002/j.1538-7305.1950.tb00463.x} {\bibfield  {journal}
  {\bibinfo  {journal} {The Bell System Technical Journal}\ }\textbf {\bibinfo
  {volume} {29}},\ \bibinfo {pages} {147} (\bibinfo {year} {1950})}\BibitemShut
  {NoStop}%
\bibitem [{\citenamefont {Derrida}\ and\ \citenamefont
  {Stauffer}(1986)}]{Derrida86}%
  \BibitemOpen
  \bibfield  {author} {\bibinfo {author} {\bibfnamefont {B.}~\bibnamefont
  {Derrida}}\ and\ \bibinfo {author} {\bibfnamefont {D.}~\bibnamefont
  {Stauffer}},\ }\bibfield  {title} {\bibinfo {title} {{Phase Transitions in
  Two-Dimensional Kauffman Cellular Automata}},\ }\href
  {https://doi.org/10.1209/0295-5075/2/10/001} {\bibfield  {journal} {\bibinfo
  {journal} {{EPL - Europhysics Letters}}\ }\textbf {\bibinfo {volume} {2}},\
  \bibinfo {pages} {739} (\bibinfo {year} {1986})}\BibitemShut {NoStop}%
\bibitem [{\citenamefont {Weisbuch}\ and\ \citenamefont
  {Stauffer}(1987)}]{Weisbuch87}%
  \BibitemOpen
  \bibfield  {author} {\bibinfo {author} {\bibfnamefont {G.}~\bibnamefont
  {Weisbuch}}\ and\ \bibinfo {author} {\bibfnamefont {D.}~\bibnamefont
  {Stauffer}},\ }\bibfield  {title} {\bibinfo {title} {{Phase transition in
  cellular random Boolean nets}},\ }\href
  {https://doi.org/10.1051/jphys:0198700480101100} {\bibfield  {journal}
  {\bibinfo  {journal} {{Journal de Physique}}\ }\textbf {\bibinfo {volume}
  {48}},\ \bibinfo {pages} {11} (\bibinfo {year} {1987})}\BibitemShut {NoStop}%
\bibitem [{\citenamefont {Wolfram}(1983)}]{Wolfram83}%
  \BibitemOpen
  \bibfield  {author} {\bibinfo {author} {\bibfnamefont {S.}~\bibnamefont
  {Wolfram}},\ }\bibfield  {title} {\bibinfo {title} {Statistical mechanics of
  cellular automata},\ }\href {https://doi.org/10.1103/RevModPhys.55.601}
  {\bibfield  {journal} {\bibinfo  {journal} {Rev. Mod. Phys.}\ }\textbf
  {\bibinfo {volume} {55}},\ \bibinfo {pages} {601} (\bibinfo {year}
  {1983})}\BibitemShut {NoStop}%
\bibitem [{\citenamefont {Alfaro}\ and\ \citenamefont
  {Sanjuán}(2024)}]{Alfaro2024}%
  \BibitemOpen
  \bibfield  {author} {\bibinfo {author} {\bibfnamefont {G.}~\bibnamefont
  {Alfaro}}\ and\ \bibinfo {author} {\bibfnamefont {M.~A.~F.}\ \bibnamefont
  {Sanjuán}},\ }\bibfield  {title} {\bibinfo {title} {Classification of
  cellular automata based on the hamming distance},\ }\href
  {https://doi.org/10.1063/5.0227349} {\bibfield  {journal} {\bibinfo
  {journal} {Chaos: An Interdisciplinary Journal of Nonlinear Science}\
  }\textbf {\bibinfo {volume} {34}},\ \bibinfo {pages} {083129} (\bibinfo
  {year} {2024})}\BibitemShut {NoStop}%
\bibitem [{\citenamefont {Buča}\ \emph {et~al.}(2021)\citenamefont {Buča},
  \citenamefont {Klobas},\ and\ \citenamefont {Prosen}}]{Buca_2021}%
  \BibitemOpen
  \bibfield  {author} {\bibinfo {author} {\bibfnamefont {B.}~\bibnamefont
  {Buča}}, \bibinfo {author} {\bibfnamefont {K.}~\bibnamefont {Klobas}},\ and\
  \bibinfo {author} {\bibfnamefont {T.}~\bibnamefont {Prosen}},\ }\bibfield
  {title} {\bibinfo {title} {Rule 54: exactly solvable model of nonequilibrium
  statistical mechanics},\ }\href {https://doi.org/10.1088/1742-5468/ac096b}
  {\bibfield  {journal} {\bibinfo  {journal} {Journal of Statistical Mechanics:
  Theory and Experiment}\ }\textbf {\bibinfo {volume} {2021}},\ \bibinfo
  {pages} {074001} (\bibinfo {year} {2021})}\BibitemShut {NoStop}%
\bibitem [{\citenamefont {Medenjak}\ \emph {et~al.}(2017)\citenamefont
  {Medenjak}, \citenamefont {Klobas},\ and\ \citenamefont
  {Prosen}}]{Medenjak17}%
  \BibitemOpen
  \bibfield  {author} {\bibinfo {author} {\bibfnamefont {M.}~\bibnamefont
  {Medenjak}}, \bibinfo {author} {\bibfnamefont {K.}~\bibnamefont {Klobas}},\
  and\ \bibinfo {author} {\bibfnamefont {T.}~\bibnamefont {Prosen}},\
  }\bibfield  {title} {\bibinfo {title} {Diffusion in deterministic interacting
  lattice systems},\ }\href {https://doi.org/10.1103/PhysRevLett.119.110603}
  {\bibfield  {journal} {\bibinfo  {journal} {Phys. Rev. Lett.}\ }\textbf
  {\bibinfo {volume} {119}},\ \bibinfo {pages} {110603} (\bibinfo {year}
  {2017})}\BibitemShut {NoStop}%
\bibitem [{\citenamefont {Prosen}\ and\ \citenamefont
  {Mejía-Monasterio}(2016)}]{Prosen_2016}%
  \BibitemOpen
  \bibfield  {author} {\bibinfo {author} {\bibfnamefont {T.}~\bibnamefont
  {Prosen}}\ and\ \bibinfo {author} {\bibfnamefont {C.}~\bibnamefont
  {Mejía-Monasterio}},\ }\bibfield  {title} {\bibinfo {title} {Integrability
  of a deterministic cellular automaton driven by stochastic boundaries},\
  }\href {https://doi.org/10.1088/1751-8113/49/18/185003} {\bibfield  {journal}
  {\bibinfo  {journal} {J. Phys. A: Math. and Theor.}\ }\textbf {\bibinfo
  {volume} {49}},\ \bibinfo {pages} {185003} (\bibinfo {year}
  {2016})}\BibitemShut {NoStop}%
\bibitem [{\citenamefont {Klobas}\ and\ \citenamefont
  {Prosen}(2022)}]{Klobas_2022}%
  \BibitemOpen
  \bibfield  {author} {\bibinfo {author} {\bibfnamefont {K.}~\bibnamefont
  {Klobas}}\ and\ \bibinfo {author} {\bibfnamefont {T.}~\bibnamefont
  {Prosen}},\ }\bibfield  {title} {\bibinfo {title} {On two reversible cellular
  automata with two particle species},\ }\href
  {https://doi.org/10.1088/1751-8121/ac3ebc} {\bibfield  {journal} {\bibinfo
  {journal} {J. of Phys. A: Math. and Theor.}\ }\textbf {\bibinfo {volume}
  {55}},\ \bibinfo {pages} {094003} (\bibinfo {year} {2022})}\BibitemShut
  {NoStop}%
\bibitem [{\citenamefont {Klobas}\ \emph {et~al.}(2021)\citenamefont {Klobas},
  \citenamefont {Bertini},\ and\ \citenamefont {Piroli}}]{Klobas21}%
  \BibitemOpen
  \bibfield  {author} {\bibinfo {author} {\bibfnamefont {K.}~\bibnamefont
  {Klobas}}, \bibinfo {author} {\bibfnamefont {B.}~\bibnamefont {Bertini}},\
  and\ \bibinfo {author} {\bibfnamefont {L.}~\bibnamefont {Piroli}},\
  }\bibfield  {title} {\bibinfo {title} {Exact thermalization dynamics in the
  ``rule 54'' quantum cellular automaton},\ }\href
  {https://doi.org/10.1103/PhysRevLett.126.160602} {\bibfield  {journal}
  {\bibinfo  {journal} {Phys. Rev. Lett.}\ }\textbf {\bibinfo {volume} {126}},\
  \bibinfo {pages} {160602} (\bibinfo {year} {2021})}\BibitemShut {NoStop}%
\bibitem [{\citenamefont {Wilkinson}\ \emph {et~al.}(2020)\citenamefont
  {Wilkinson}, \citenamefont {Klobas}, \citenamefont {Prosen},\ and\
  \citenamefont {Garrahan}}]{Wilkinson20}%
  \BibitemOpen
  \bibfield  {author} {\bibinfo {author} {\bibfnamefont {J.~W.~P.}\
  \bibnamefont {Wilkinson}}, \bibinfo {author} {\bibfnamefont {K.}~\bibnamefont
  {Klobas}}, \bibinfo {author} {\bibfnamefont {T.}~\bibnamefont {Prosen}},\
  and\ \bibinfo {author} {\bibfnamefont {J.~P.}\ \bibnamefont {Garrahan}},\
  }\bibfield  {title} {\bibinfo {title} {Exact solution of the floquet-pxp
  cellular automaton},\ }\href {https://doi.org/10.1103/PhysRevE.102.062107}
  {\bibfield  {journal} {\bibinfo  {journal} {Phys. Rev. E}\ }\textbf {\bibinfo
  {volume} {102}},\ \bibinfo {pages} {062107} (\bibinfo {year}
  {2020})}\BibitemShut {NoStop}%
\bibitem [{\citenamefont {Gombor}\ and\ \citenamefont
  {Pozsgay}(2022)}]{Pozsgay}%
  \BibitemOpen
  \bibfield  {author} {\bibinfo {author} {\bibfnamefont {T.}~\bibnamefont
  {Gombor}}\ and\ \bibinfo {author} {\bibfnamefont {B.}~\bibnamefont
  {Pozsgay}},\ }\bibfield  {title} {\bibinfo {title} {{Superintegrable cellular
  automata and dual unitary gates from Yang-Baxter maps}},\ }\href
  {https://doi.org/10.21468/SciPostPhys.12.3.102} {\bibfield  {journal}
  {\bibinfo  {journal} {SciPost Phys.}\ }\textbf {\bibinfo {volume} {12}},\
  \bibinfo {pages} {102} (\bibinfo {year} {2022})}\BibitemShut {NoStop}%
\bibitem [{\citenamefont {Prosen}(2023)}]{Prosen2023}%
  \BibitemOpen
  \bibfield  {author} {\bibinfo {author} {\bibfnamefont {T.}~\bibnamefont
  {Prosen}},\ }\bibfield  {title} {\bibinfo {title} {On two non-ergodic
  reversible cellular automata, one classical, the other quantum},\ }\href
  {https://doi.org/10.3390/e25050739} {\bibfield  {journal} {\bibinfo
  {journal} {Entropy}\ }\textbf {\bibinfo {volume} {25}},\ \bibinfo {pages}
  {739} (\bibinfo {year} {2023})}\BibitemShut {NoStop}%
\bibitem [{\citenamefont {Gombor}\ and\ \citenamefont
  {Pozsgay}(2024)}]{Gombor2024}%
  \BibitemOpen
  \bibfield  {author} {\bibinfo {author} {\bibfnamefont {T.}~\bibnamefont
  {Gombor}}\ and\ \bibinfo {author} {\bibfnamefont {B.}~\bibnamefont
  {Pozsgay}},\ }\bibfield  {title} {\bibinfo {title} {{Integrable deformations
  of superintegrable quantum circuits}},\ }\href
  {https://doi.org/10.21468/SciPostPhys.16.4.114} {\bibfield  {journal}
  {\bibinfo  {journal} {SciPost Phys.}\ }\textbf {\bibinfo {volume} {16}},\
  \bibinfo {pages} {114} (\bibinfo {year} {2024})}\BibitemShut {NoStop}%
\bibitem [{\citenamefont {Sharipov}\ \emph {et~al.}(2025)\citenamefont
  {Sharipov}, \citenamefont {Koterle}, \citenamefont {Grozdanov},\ and\
  \citenamefont {Prosen}}]{Rustem}%
  \BibitemOpen
  \bibfield  {author} {\bibinfo {author} {\bibfnamefont {R.}~\bibnamefont
  {Sharipov}}, \bibinfo {author} {\bibfnamefont {M.}~\bibnamefont {Koterle}},
  \bibinfo {author} {\bibfnamefont {S.}~\bibnamefont {Grozdanov}},\ and\
  \bibinfo {author} {\bibfnamefont {T.}~\bibnamefont {Prosen}},\ }\href
  {https://arxiv.org/abs/2503.16593} {\bibinfo {title} {Ergodic behaviors in
  reversible 3-state cellular automata}} (\bibinfo {year} {2025}),\ \Eprint
  {https://arxiv.org/abs/2503.16593} {arXiv:2503.16593 [cond-mat.stat-mech]}
  \BibitemShut {NoStop}%
\bibitem [{\citenamefont {Krajnik}\ \emph {et~al.}(2022)\citenamefont
  {Krajnik}, \citenamefont {Schmidt}, \citenamefont {Pasquier}, \citenamefont
  {Ilievski},\ and\ \citenamefont {Prosen}}]{Krajnik22}%
  \BibitemOpen
  \bibfield  {author} {\bibinfo {author} {\bibfnamefont {{\v{Z}}.}~\bibnamefont
  {Krajnik}}, \bibinfo {author} {\bibfnamefont {J.}~\bibnamefont {Schmidt}},
  \bibinfo {author} {\bibfnamefont {V.}~\bibnamefont {Pasquier}}, \bibinfo
  {author} {\bibfnamefont {E.}~\bibnamefont {Ilievski}},\ and\ \bibinfo
  {author} {\bibfnamefont {T.}~\bibnamefont {Prosen}},\ }\bibfield  {title}
  {\bibinfo {title} {Exact anomalous current fluctuations in a deterministic
  interacting model},\ }\href {https://doi.org/10.1103/PhysRevLett.128.160601}
  {\bibfield  {journal} {\bibinfo  {journal} {Phys. Rev. Lett.}\ }\textbf
  {\bibinfo {volume} {128}},\ \bibinfo {pages} {160601} (\bibinfo {year}
  {2022})}\BibitemShut {NoStop}%
\bibitem [{\citenamefont {Krajnik}\ \emph {et~al.}(2024)\citenamefont
  {Krajnik}, \citenamefont {Schmidt}, \citenamefont {Pasquier}, \citenamefont
  {Prosen},\ and\ \citenamefont {Ilievski}}]{Krajnik24}%
  \BibitemOpen
  \bibfield  {author} {\bibinfo {author} {\bibfnamefont {{\v{Z}}.}~\bibnamefont
  {Krajnik}}, \bibinfo {author} {\bibfnamefont {J.}~\bibnamefont {Schmidt}},
  \bibinfo {author} {\bibfnamefont {V.}~\bibnamefont {Pasquier}}, \bibinfo
  {author} {\bibfnamefont {T.}~\bibnamefont {Prosen}},\ and\ \bibinfo {author}
  {\bibfnamefont {E.}~\bibnamefont {Ilievski}},\ }\bibfield  {title} {\bibinfo
  {title} {Universal anomalous fluctuations in charged single-file systems},\
  }\href {https://doi.org/10.1103/PhysRevResearch.6.013260} {\bibfield
  {journal} {\bibinfo  {journal} {Phys. Rev. Res.}\ }\textbf {\bibinfo {volume}
  {6}},\ \bibinfo {pages} {013260} (\bibinfo {year} {2024})}\BibitemShut
  {NoStop}%
\bibitem [{\citenamefont {Krajnik}\ \emph {et~al.}(2025)\citenamefont
  {Krajnik}, \citenamefont {Klobas}, \citenamefont {Bertini},\ and\
  \citenamefont {Prosen}}]{Krajnik25}%
  \BibitemOpen
  \bibfield  {author} {\bibinfo {author} {\bibfnamefont {{\v{Z}}.}~\bibnamefont
  {Krajnik}}, \bibinfo {author} {\bibfnamefont {K.}~\bibnamefont {Klobas}},
  \bibinfo {author} {\bibfnamefont {B.}~\bibnamefont {Bertini}},\ and\ \bibinfo
  {author} {\bibfnamefont {T.}~\bibnamefont {Prosen}},\ }\bibfield  {title}
  {\bibinfo {title} {Fluctuations of stochastic charged cellular automata},\
  }\href {https://doi.org/10.1088/1742-5468/add513} {\bibfield  {journal}
  {\bibinfo  {journal} {Journal of Statistical Mechanics: Theory and
  Experiment}\ }\textbf {\bibinfo {volume} {2025}},\ \bibinfo {pages} {053209}
  (\bibinfo {year} {2025})}\BibitemShut {NoStop}%
\bibitem [{\citenamefont {Gopalakrishnan}(2018)}]{Gopalakrishnan2018}%
  \BibitemOpen
  \bibfield  {author} {\bibinfo {author} {\bibfnamefont {S.}~\bibnamefont
  {Gopalakrishnan}},\ }\bibfield  {title} {\bibinfo {title} {Operator growth
  and eigenstate entanglement in an interacting integrable floquet system},\
  }\href {https://doi.org/10.1103/PhysRevB.98.060302} {\bibfield  {journal}
  {\bibinfo  {journal} {Phys. Rev. B}\ }\textbf {\bibinfo {volume} {98}},\
  \bibinfo {pages} {060302} (\bibinfo {year} {2018})}\BibitemShut {NoStop}%
\bibitem [{\citenamefont {Gopalakrishnan}\ \emph {et~al.}(2018)\citenamefont
  {Gopalakrishnan}, \citenamefont {Huse}, \citenamefont {Khemani},\ and\
  \citenamefont {Vasseur}}]{Gopalakrishnan2018_2}%
  \BibitemOpen
  \bibfield  {author} {\bibinfo {author} {\bibfnamefont {S.}~\bibnamefont
  {Gopalakrishnan}}, \bibinfo {author} {\bibfnamefont {D.~A.}\ \bibnamefont
  {Huse}}, \bibinfo {author} {\bibfnamefont {V.}~\bibnamefont {Khemani}},\ and\
  \bibinfo {author} {\bibfnamefont {R.}~\bibnamefont {Vasseur}},\ }\bibfield
  {title} {\bibinfo {title} {Hydrodynamics of operator spreading and
  quasiparticle diffusion in interacting integrable systems},\ }\href
  {https://doi.org/10.1103/PhysRevB.98.220303} {\bibfield  {journal} {\bibinfo
  {journal} {Phys. Rev. B}\ }\textbf {\bibinfo {volume} {98}},\ \bibinfo
  {pages} {220303} (\bibinfo {year} {2018})}\BibitemShut {NoStop}%
\bibitem [{\citenamefont {Alba}\ \emph {et~al.}(2019)\citenamefont {Alba},
  \citenamefont {Dubail},\ and\ \citenamefont {Medenjak}}]{Alba2019}%
  \BibitemOpen
  \bibfield  {author} {\bibinfo {author} {\bibfnamefont {V.}~\bibnamefont
  {Alba}}, \bibinfo {author} {\bibfnamefont {J.}~\bibnamefont {Dubail}},\ and\
  \bibinfo {author} {\bibfnamefont {M.}~\bibnamefont {Medenjak}},\ }\bibfield
  {title} {\bibinfo {title} {Operator entanglement in interacting integrable
  quantum systems: The case of the rule 54 chain},\ }\href
  {https://doi.org/10.1103/PhysRevLett.122.250603} {\bibfield  {journal}
  {\bibinfo  {journal} {Phys. Rev. Lett.}\ }\textbf {\bibinfo {volume} {122}},\
  \bibinfo {pages} {250603} (\bibinfo {year} {2019})}\BibitemShut {NoStop}%
\bibitem [{\citenamefont {Klobas}(2024)}]{Klobas_2024}%
  \BibitemOpen
  \bibfield  {author} {\bibinfo {author} {\bibfnamefont {K.}~\bibnamefont
  {Klobas}},\ }\bibfield  {title} {\bibinfo {title} {Non-equilibrium dynamics
  of symmetry-resolved entanglement and entanglement asymmetry: exact
  asymptotics in rule 54*},\ }\href {https://doi.org/10.1088/1751-8121/ad91fd}
  {\bibfield  {journal} {\bibinfo  {journal} {Journal of Physics A:
  Mathematical and Theoretical}\ }\textbf {\bibinfo {volume} {57}},\ \bibinfo
  {pages} {505001} (\bibinfo {year} {2024})}\BibitemShut {NoStop}%
\bibitem [{\citenamefont {De~Fazio}\ \emph {et~al.}(2024)\citenamefont
  {De~Fazio}, \citenamefont {Garrahan},\ and\ \citenamefont
  {Klobas}}]{De_Fazio_2024}%
  \BibitemOpen
  \bibfield  {author} {\bibinfo {author} {\bibfnamefont {C.}~\bibnamefont
  {De~Fazio}}, \bibinfo {author} {\bibfnamefont {J.~P.}\ \bibnamefont
  {Garrahan}},\ and\ \bibinfo {author} {\bibfnamefont {K.}~\bibnamefont
  {Klobas}},\ }\bibfield  {title} {\bibinfo {title} {Exact results on the
  dynamics of the stochastic floquet-east model*},\ }\href
  {https://doi.org/10.1088/1751-8121/ad8e1c} {\bibfield  {journal} {\bibinfo
  {journal} {Journal of Physics A: Mathematical and Theoretical}\ }\textbf
  {\bibinfo {volume} {57}},\ \bibinfo {pages} {505002} (\bibinfo {year}
  {2024})}\BibitemShut {NoStop}%
\bibitem [{\citenamefont {Klobas}\ \emph {et~al.}(2024)\citenamefont {Klobas},
  \citenamefont {De~Fazio},\ and\ \citenamefont {Garrahan}}]{Klobas2024_2}%
  \BibitemOpen
  \bibfield  {author} {\bibinfo {author} {\bibfnamefont {K.}~\bibnamefont
  {Klobas}}, \bibinfo {author} {\bibfnamefont {C.}~\bibnamefont {De~Fazio}},\
  and\ \bibinfo {author} {\bibfnamefont {J.~P.}\ \bibnamefont {Garrahan}},\
  }\bibfield  {title} {\bibinfo {title} {Exact pretransition effects in
  kinetically constrained circuits: Dynamical fluctuations in the floquet-east
  model},\ }\href {https://doi.org/10.1103/PhysRevE.110.L022101} {\bibfield
  {journal} {\bibinfo  {journal} {Phys. Rev. E}\ }\textbf {\bibinfo {volume}
  {110}},\ \bibinfo {pages} {L022101} (\bibinfo {year} {2024})}\BibitemShut
  {NoStop}%
\bibitem [{\citenamefont {Sala}\ \emph {et~al.}(2020)\citenamefont {Sala},
  \citenamefont {Rakovszky}, \citenamefont {Verresen}, \citenamefont {Knap},\
  and\ \citenamefont {Pollmann}}]{Pollmann}%
  \BibitemOpen
  \bibfield  {author} {\bibinfo {author} {\bibfnamefont {P.}~\bibnamefont
  {Sala}}, \bibinfo {author} {\bibfnamefont {T.}~\bibnamefont {Rakovszky}},
  \bibinfo {author} {\bibfnamefont {R.}~\bibnamefont {Verresen}}, \bibinfo
  {author} {\bibfnamefont {M.}~\bibnamefont {Knap}},\ and\ \bibinfo {author}
  {\bibfnamefont {F.}~\bibnamefont {Pollmann}},\ }\bibfield  {title} {\bibinfo
  {title} {Ergodicity breaking arising from hilbert space fragmentation in
  dipole-conserving hamiltonians},\ }\href
  {https://doi.org/10.1103/PhysRevX.10.011047} {\bibfield  {journal} {\bibinfo
  {journal} {Phys. Rev. X}\ }\textbf {\bibinfo {volume} {10}},\ \bibinfo
  {pages} {011047} (\bibinfo {year} {2020})}\BibitemShut {NoStop}%
\bibitem [{\citenamefont {Moudgalya}\ \emph {et~al.}(2022)\citenamefont
  {Moudgalya}, \citenamefont {Bernevig},\ and\ \citenamefont
  {Regnault}}]{Moudgalya_2022}%
  \BibitemOpen
  \bibfield  {author} {\bibinfo {author} {\bibfnamefont {S.}~\bibnamefont
  {Moudgalya}}, \bibinfo {author} {\bibfnamefont {B.~A.}\ \bibnamefont
  {Bernevig}},\ and\ \bibinfo {author} {\bibfnamefont {N.}~\bibnamefont
  {Regnault}},\ }\bibfield  {title} {\bibinfo {title} {Quantum many-body scars
  and hilbert space fragmentation: a review of exact results},\ }\href
  {https://doi.org/10.1088/1361-6633/ac73a0} {\bibfield  {journal} {\bibinfo
  {journal} {Reports on Progress in Physics}\ }\textbf {\bibinfo {volume}
  {85}},\ \bibinfo {pages} {086501} (\bibinfo {year} {2022})}\BibitemShut
  {NoStop}%
\bibitem [{\citenamefont {Serbyn}\ \emph {et~al.}(2013)\citenamefont {Serbyn},
  \citenamefont {Papi{\'{c}}},\ and\ \citenamefont {Abanin}}]{Serbyn2013}%
  \BibitemOpen
  \bibfield  {author} {\bibinfo {author} {\bibfnamefont {M.}~\bibnamefont
  {Serbyn}}, \bibinfo {author} {\bibfnamefont {Z.}~\bibnamefont
  {Papi{\'{c}}}},\ and\ \bibinfo {author} {\bibfnamefont {D.~A.}\ \bibnamefont
  {Abanin}},\ }\bibfield  {title} {\bibinfo {title} {Local conservation laws
  and the structure of the many-body localized states},\ }\href
  {https://doi.org/10.1103/PhysRevLett.111.127201} {\bibfield  {journal}
  {\bibinfo  {journal} {Phys. Rev. Lett.}\ }\textbf {\bibinfo {volume} {111}},\
  \bibinfo {pages} {127201} (\bibinfo {year} {2013})}\BibitemShut {NoStop}%
\bibitem [{\citenamefont {Abanin}\ \emph {et~al.}(2019)\citenamefont {Abanin},
  \citenamefont {Altman}, \citenamefont {Bloch},\ and\ \citenamefont
  {Serbyn}}]{Abanin19}%
  \BibitemOpen
  \bibfield  {author} {\bibinfo {author} {\bibfnamefont {D.~A.}\ \bibnamefont
  {Abanin}}, \bibinfo {author} {\bibfnamefont {E.}~\bibnamefont {Altman}},
  \bibinfo {author} {\bibfnamefont {I.}~\bibnamefont {Bloch}},\ and\ \bibinfo
  {author} {\bibfnamefont {M.}~\bibnamefont {Serbyn}},\ }\bibfield  {title}
  {\bibinfo {title} {Colloquium: Many-body localization, thermalization, and
  entanglement},\ }\href {https://doi.org/10.1103/RevModPhys.91.021001}
  {\bibfield  {journal} {\bibinfo  {journal} {Rev. Mod. Phys.}\ }\textbf
  {\bibinfo {volume} {91}},\ \bibinfo {pages} {021001} (\bibinfo {year}
  {2019})}\BibitemShut {NoStop}%
\bibitem [{\citenamefont {Sierant}\ \emph {et~al.}(2025)\citenamefont
  {Sierant}, \citenamefont {Lewenstein}, \citenamefont {Scardicchio},
  \citenamefont {Vidmar},\ and\ \citenamefont {Zakrzewski}}]{Sierant_2025}%
  \BibitemOpen
  \bibfield  {author} {\bibinfo {author} {\bibfnamefont {P.}~\bibnamefont
  {Sierant}}, \bibinfo {author} {\bibfnamefont {M.}~\bibnamefont {Lewenstein}},
  \bibinfo {author} {\bibfnamefont {A.}~\bibnamefont {Scardicchio}}, \bibinfo
  {author} {\bibfnamefont {L.}~\bibnamefont {Vidmar}},\ and\ \bibinfo {author}
  {\bibfnamefont {J.}~\bibnamefont {Zakrzewski}},\ }\bibfield  {title}
  {\bibinfo {title} {Many-body localization in the age of classical
  computing*},\ }\href {https://doi.org/10.1088/1361-6633/ad9756} {\bibfield
  {journal} {\bibinfo  {journal} {Reports on Progress in Physics}\ }\textbf
  {\bibinfo {volume} {88}},\ \bibinfo {pages} {026502} (\bibinfo {year}
  {2025})}\BibitemShut {NoStop}%
\bibitem [{Note1()}]{Note1}%
  \BibitemOpen
  \bibinfo {note} {The local map $\Phi _v$ can as well be identified with a
  $2^4\times 2^4$ (permutation) matrix which is embedded in $2^{N_{\protect \rm
  E}}$ dimensional Hilbert space $\protect \mathbb C^{2^{N_{\protect \rm E}}}$
  of $N_{\protect \rm E}$ qubits. Similarly, the complete dynamical map $\Phi $
  is a $2^{N_{\protect \rm E}} \times 2^{N_{\protect \rm E}}$ matrix
  representing an element in $S(2^{N_{\protect \rm E}})$, i.e. deterministic
  dynamics. Let us define observables as functions over phase space $a :
  \protect \mathbb Z_2^{N_{\protect \rm E}} \to \protect \mathbb C$, or
  equivalently as vectors in $\protect \mathbb C^{2^{N_{\protect \rm E}}}$. The
  observable $q$ is a conserved quantity if it satisfies $q\circ \Phi = q$, or
  $q \Phi = q$ in vector notation.}\BibitemShut {Stop}%
\bibitem [{\citenamefont {Orlov}\ \emph {et~al.}()\citenamefont {Orlov}, ,
  \citenamefont {Jonay},\ and\ \citenamefont {Prosen}}]{PavelOrlov}%
  \BibitemOpen
  \bibfield  {author} {\bibinfo {author} {\bibfnamefont {P.}~\bibnamefont
  {Orlov}}, , \bibinfo {author} {\bibfnamefont {C.}~\bibnamefont {Jonay}},\
  and\ \bibinfo {author} {\bibfnamefont {T.}~\bibnamefont {Prosen}},\ }\bibinfo
  {title} {To be published}\BibitemShut {NoStop}%
\bibitem [{Note2()}]{Note2}%
  \BibitemOpen
\bibfield  {title} {  }\bibinfo {note} {In contrast with topological loop
  charges, which act on $N$ degrees of freedom, and so, the density $M_{\gamma
  }/N$ for these charges can be well-defined in the thermodynamic
  limit}\BibitemShut {NoStop}%
\end{thebibliography}%

\end{document}